\documentclass[sigconf]{acmart}

\AtBeginDocument{%
  }

\setcopyright{acmlicensed}
\copyrightyear{2026}
\acmYear{2026}
\setcopyright{cc}
\setcctype{by}
\acmConference[CHI '26]{Proceedings of the 2026 CHI Conference on Human Factors in Computing Systems}{April 13--17, 2026}{Barcelona, Spain}
\acmBooktitle{Proceedings of the 2026 CHI Conference on Human Factors in Computing Systems (CHI '26), April 13--17, 2026, Barcelona, Spain}
\acmPrice{}
\acmDOI{10.1145/3772318.3790958}
\acmISBN{979-8-4007-2278-3/2026/04}

\usepackage{circledsteps}
\usepackage{ifthen}
\usepackage{xcolor}
\usepackage[normalem]{ulem} 
\begin{document}

\newcommand{\dora}[1]{\textcolor{blue}{[dora: #1]}}
\newcommand{\diyi}[1]{\textcolor{red}{[diyi: #1]}}
\newcommand{\hannah}[1]{\textcolor{orange}{[hannah: #1]}}
\newcommand{\new}[1]{\textcolor{black}{#1}}
\newcommand{\michael}[1]{\textcolor{cyan}{[michael: #1]}}

\newboolean{markup}
\setboolean{markup}{false}   

\newcommand{\add}[1]{%
  \ifthenelse{\boolean{markup}}
    {\textcolor{blue}{#1}}
    {#1}
}

\newcommand{\addtable}[1]{%
  \ifthenelse{\boolean{markup}}
    {{\color{blue}#1}}
    {#1}
}

\newcommand{\remove}[1]{%
  \ifthenelse{\boolean{markup}}
    {{\color{red}\sout{#1}}}
    {}
}
\raggedbottom
\title{Whose Knowledge Counts? Co-Designing Community-Centered AI Auditing Tools with Educators in Hawai`i}

\author{Dora Zhao}
\affiliation{%
  \institution{Stanford University}
  \city{Stanford}
  \state{CA}
  \country{USA}}
\email{dorothyz@stanford.edu}

\author{Hannah Cha}
\affiliation{%
  \institution{Stanford University}
  \city{Stanford}
  \state{CA}
  \country{USA}}
\email{hcha417@stanford.edu}

\author{Michael J. Ryan}
\affiliation{%
  \institution{Stanford University}
  \city{Stanford}
  \state{CA}
  \country{USA}}
\email{michaeljryan@stanford.edu}

\author{Angelina Wang}
\affiliation{%
  \institution{Cornell Tech}
  \city{New York}
  \state{NY}
  \country{USA}}
\email{angelina.wang@cornell.edu}

\author{Rachel Baker-Ramos}
\affiliation{%
  \institution{Georgia Institute of Technology}
  \city{Atlanta}
  \state{GA}
  \country{USA}}
\email{rachelbaker@gatech.edu}

\author{Evyn-Bree Helekahi-Kaiwi}
\affiliation{%
  \institution{Ulu Lāhui Foundation}
  \city{Honolulu}
  \state{HI}
  \country{USA}}
\email{evyn-bree.hele-kaiwi@ululahui.org}

\author{Rebecca Diego}
\affiliation{%
  \institution{Ulu Lāhui Foundation}
  \city{Honolulu}
  \state{HI}
  \country{USA}}
\email{rebecca.diego@ululahui.org}

\author{Josiah Hester}
\affiliation{%
  \institution{Georgia Institute of Technology}
  \city{Atlanta}
  \state{GA}
  \country{USA}}
\email{josiah@gatech.edu}

\author{Diyi Yang}
\affiliation{%
  \institution{Stanford University}
  \city{Stanford}
  \state{CA}
  \country{USA}}
\email{diyiy@cs.stanford.edu}

\renewcommand{\shortauthors}{Zhao et al.}

\begin{abstract}
Although generative AI is being deployed into classrooms with promises of aiding teachers, educators caution that these tools can have unintended pedagogical repercussions, including cultural misrepresentation and bias. These concerns are heightened in low-resource language and Indigenous education settings, where AI systems frequently underperform. We investigate these challenges in Hawai`i, where public schools operate under a statewide mandate to integrate Hawaiian language and culture into education. Through four co-design workshops with 22 public school educators, we surfaced concerns about using generative AI in educational settings, particularly around cultural misrepresentation, and corresponding designs for auditing tools that address these issues. We find that educators envision tools grounded in specific Hawaiian cultural values and practices, such as tracing the genealogy of knowledge in source materials. Building on these insights, we conceptualize AI auditing as a \emph{community-oriented process} rather than the work of isolated individuals, and discuss implications for designing auditing tools.
\end{abstract}



\maketitle
\section{Introduction}
\begin{figure*}
\includegraphics[width=\linewidth]{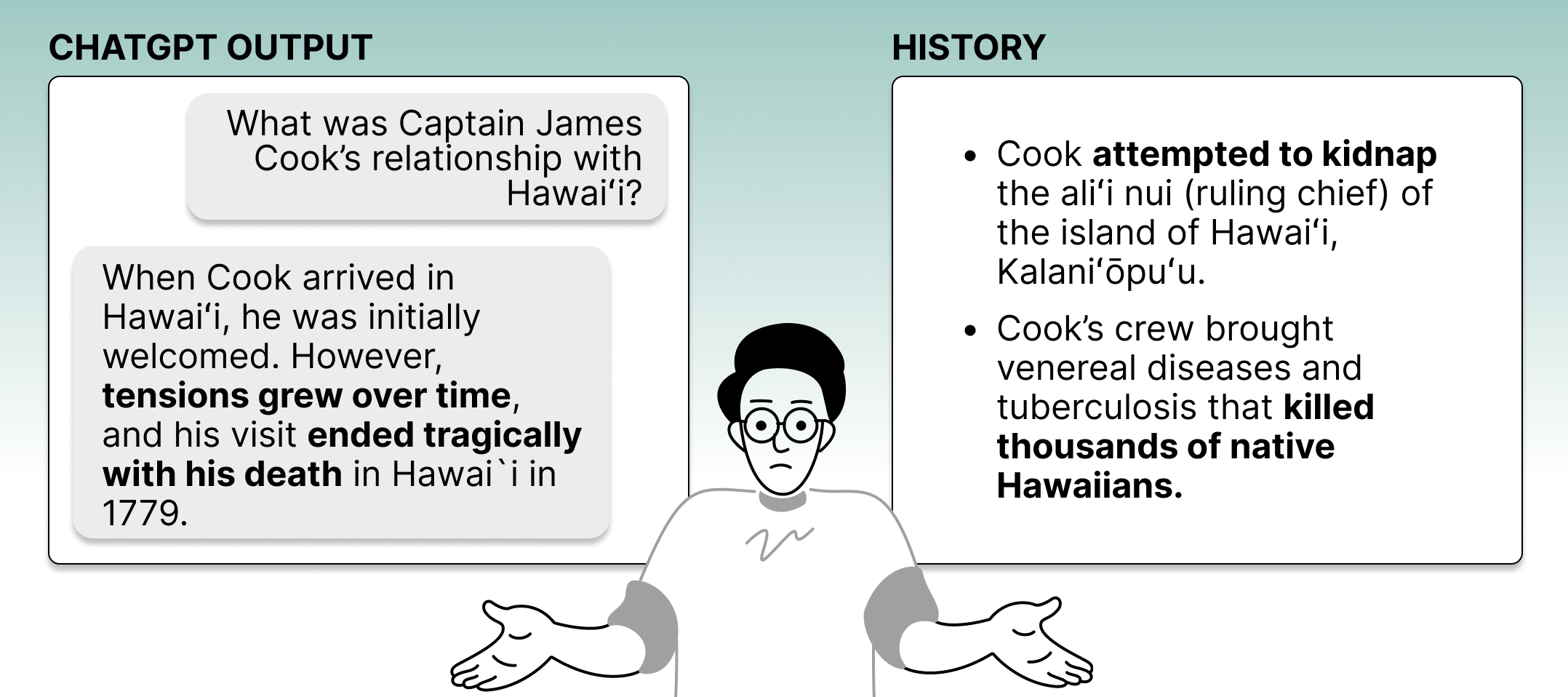}
  \caption{Output from ChatGPT when queried about Captain James Cook's relationship with Hawai`i presents a more sanitized depiction of Cook, failing to highlight the negative impacts he had on Hawai`i~\cite{history2025cookdeath, salmond2004trial}.}
  \label{fig:teaser}
  \Description[A side-by-side comparison shows the discrepancy between ChatGPT’s response and historical accounts of Captain James Cook’s relationship with Hawai`i.]{A side-by-side comparison shows the discrepancy between ChatGPT’s response and historical accounts of Captain James Cook’s relationship with Hawai`i. On the left, the ChatGPT output presents a simplified summary: Cook was initially welcomed, tensions grew, and his visit ended tragically with his death in 1779. On the right, the historical record highlights more serious harms: Cook attempted to kidnap Kalani`ōpu`u, the ruling chief of Hawai`i Island, and his crew introduced diseases that killed thousands of Native Hawaiians.}
\end{figure*}

Generative AI systems, such as large language models (LLMs), are being increasingly deployed for educational purposes. These technologies are touted to increase educators' capacity to support students while maintaining high-quality instruction~\cite{hashem2024ai} and provide inexpensive ways to cater to diverse student populations~\cite{cardona2023ai, dai2023can, meyer2024using}. Already, teachers have adopted generative AI to assist with many different tasks, including automated lesson planning~\cite{kehoe2023leveraging, baytak2024content, fan2024lessonplanner, karaman2024lesson}, brainstorming teaching strategies~\cite{kehoe2023leveraging, mogavi2024chatgpt, hu2024teaching, lee2024using}, and adapting learning to increase student engagement~\cite{alier2023smart, sharma2024comuniqa}. However, these technological solutions do not come without concerns: generative AI models are known to reproduce or exacerbate social and cultural biases by misrepresenting cultural content and overrepresenting Western narratives~\cite{liu2025cultural, tao2024cultural}. For instance, when prompted with controversial questions, Anthropic's Claude answers 11.7\% more similarly to U.S. perspectives than Chinese ones \cite{durmus2023towards}. This bias manifests in educational contexts as well. As shown in Fig. \ref{fig:teaser}, generated outputs perpetuate dominant Western narratives, glorifying Captain Cook's relationship with Hawai`i and glossing over the known harms he caused~\cite{history2025cookdeath, salmond2004trial}. In doing so, these outputs reproduce imperial legacies, obscure the profound repercussions on Indigenous communities, and skew understandings of histories.

These misrepresentations have outsized consequences when AI systems are deployed in education settings as these technologies can shape both \emph{what} students learn and \emph{how} they learn, such as impacting critical-thinking abilities or social development~\cite{harvey2025don}. As is the case for the example in Fig.~\ref{fig:teaser}, this risk is compounded in the context of low-resource languages and Indigenous education settings, where AI systems are more prone to inaccuracies and mistranslations~\cite{pinhanez2023balancing, zhong2024opportunitieschallengeslargelanguage, hasan2024large}. However, while teachers want to be able to assess the quality and cultural relevance of AI-generated outputs, they often lack the time, tools, or training to make this tractable~\cite{cabellos2024university, kasneci2023chatgpt, moura2024teachers, mhasakar2025itrustwesternkumu, pappa2024technology}.

One hope is that model providers can proactively intervene to address these concerns. In practice, however, many institutional barriers hinder the adoption of responsible AI practices, making top-down approaches unreliable~\cite{Varanasi_2023,holstein2019improving}. A fruitful alternative are bottom-up efforts in which end users act as auditors, surfacing harmful algorithmic behaviors they encounter in their day-to-day interactions~\cite{shen2021everyday,devos2022toward}. Since end users possess situated knowledge about how tools are used in practice, they can identify harms practitioners or outsiders would overlook~\cite{mack2024they,mim2024between,shelby2024generative,zhang2024partiality}. Yet, many end users lack the technical expertise to conduct systematic audits independently, highlighting the need for auditing tools designed to scaffold their efforts~\cite{lam2022end,deng2025weauditscaffoldinguserauditors}. While prior end-user auditing systems have been proposed for more general settings, educational contexts introduce additional challenges: K–12 teachers face high-stakes environments, understaffed classrooms~\cite{nguyen2024we}, and overwhelming workloads~\cite{geiger2018effects}, alongside culturally situated demands to ensure AI-generated outputs align with curricular standards and pedagogical practices. These conditions raise critical questions about how auditing practices can be effectively integrated into educational contexts, such as what concerns audits should prioritize and what tools are best suited to support teachers.

To make progress on answering these questions, we conduct a case study on co-designing AI auditing tools with public school educators in O`ahu, Hawai`i. We situate our study in Hawai`i due to the unique intersection of educational and cultural challenges that arise when integrating generative AI systems in this setting. Public school educators in Hawai`i are rapidly adopting AI systems for pedagogical purposes~\cite{hawaiiAI2025}. At the same time, public schools in Hawai`i operate under a government mandate to incorporate Hawaiian language and culture~\cite{ohe2015plan, LRB_Hawaii_ConConStudies1978}, and Hawai`i is one of the few states to have public language immersion programs in an Indigenous language~\cite{hoffman2023advancing}, \textit{`Ōlelo Hawai`i}~\cite{pinhanez2023balancing}. The dual pressure of harnessing generative AI for pedagogical benefit while avoiding cultural inaccuracies have already created tension among educators~\cite{mhasakar2025itrustwesternkumu}. \new{By focusing on this intersection, our case study surfaces how educators navigate both pedagogical and sociotechnical concerns regarding generative AI in culturally sensitive contexts.}

In this work, we conducted four co-design workshops with 22 public school educators. During the workshops, participants reflected on their current uses of generative AI, discussed instances of cultural misrepresentation they encountered, and engaged in design exercises to ideate potential auditing tools. The proposed tools centered on three priorities: identifying sources for generated outputs, distinguishing whether Western narratives or Kanaka (Native Hawaiian) experiences were represented, and flagging problematic model outputs. While these features resemble those supported by general-purpose auditing tools, our findings uncover the nuances behind participants' requests that were shaped by community values and practices, impacting the resulting designs. For example, unlike conventional attribution or retrieval-based approaches~\cite{chen2024honestaifinetuningsmall, mohammed2025aftina, njeh2024enhancing}, participants emphasized the importance of tracing the \emph{genealogy} of sources given its central role in Hawaiian cultural practice. 

\new{Our findings on the specific harms and designs surfaced in our workshops are tied to a select group of educators in O`ahu and may not generalize to other settings. However, our broader insight --- that general-purpose auditing processes often fail to accommodate local epistemologies, values, and knowledge systems --- extends beyond this context. Drawing from decolonial perspectives within HCI~\cite{alvarado2021decolonial,ali2016brief,wong2020decolonizing}, we argue that designing effective auditing practices requires centering the plurality of community-specific ways of knowing rather than assuming a universal, Western default.} Under this framing, we argue for reorienting end-user auditing as a \emph{community-oriented process} rather than a task carried out by isolated individuals, and propose \new{three design recommendations:~(1) determining who should participate as auditors,~(2) developing infrastructures that embed community values, and~(3) reorienting the desired outcomes of audits.}

In total, we contribute the following:
\begin{itemize}
\item \textbf{Dimensions of cultural misrepresentation and corresponding auditing tool designs}: We provide empirical insights from our co-design workshops on the types of cultural misrepresentations educators encounter when using generative AI and highlight auditing tools they proposed to address these concerns.

\item \textbf{Guidelines for designing community-centered auditing tools and infrastructures}: We argue for framing auditing as a \emph{community-oriented process}. Using decoloniality as a lens, we discuss design implications related to supporting diverse community perspectives, involving community members in the creation of auditing tools, and reconsidering the intended outcomes for AI auditing. 
\end{itemize}

\section{Background}
\label{sec:background}

In this section, we provide background on Hawai`i and its public education system to contextualize the relevance and importance of this study.

\subsection{Social and Cultural Context: Hawai`i}
We situate our study in Hawai`i, as it presents a unique context where factors shaping culturally grounded AI auditing are critically converging in the education domain. Hawai`i is home to \emph{Kanaka Maoli} (Native Hawaiians) and is the only U.S. state to designate a Native language, \emph{`Ōlelo Hawai`i}, as an official state language~\cite{NPS_OleloHawaii_2023}. Beyond its strong Indigenous presence, Hawai`i is also among the most multicultural states in the U.S., with the highest proportion of Multiracial residents nationwide~\cite{uscensus_hawaii_state_local_geo_guides_2010}. Hawaiian researchers have noted that island systems can also serve as model systems for studying social and technical interventions given the bounded environment~\cite{beamer2023island}. Recognizing that community-based research can unintentionally exploit or harm communities, even when well-intentioned~\cite{harrington2019deconstructing,gugganig2021hawai`i}, our work is grounded in a commitment to centering the needs of the communities we engage with. In this context, Hawai`i offers a paradoxical advantage: despite the complexity of its cultural landscape, its island scale enables interventions to propagate more rapidly and be observed more clearly than in larger, more diffuse systems~\cite{beamer2023island}.

\subsection{Cultural Education and AI Usage in Hawai`i's Public Education System}
The education system in Hawai`i represents a unique confluence where new generative AI systems are being utilized for curriculum that cover culturally sensitive topics. Public schools in Hawaii are rapidly adopting AI technologies, spurred by state-level partnerships with AI providers~\cite{hawaiiAI2025}. This rapid integration intersects with long-standing commitments to culturally relevant education in Hawai`i. Since 1987, the Department of Education in Hawai`i has a mandate to integrate the study of Hawaiian culture and language in classrooms; this mandate was further bolstered in 2015 when the Board of Education instituted policies around integrating Hawaiian education into classroom standards~\cite{ohe2015plan}. The Department of Education acknowledges that \textit{``the knowledge of our kūpuna is the guiding light that directs our purpose in support of Hawaiian education''}, highlighting that learning must reflect Indigenous teachings and values~\cite{NPS_OleloHawaii_2023}. As a part of these cultural revitalization efforts, 22 public schools in Hawai`i offer \emph{Ka Papahana Kaiapuni Hawai`i}, or Hawaiian language immersion programs, which contain culturally grounded instruction taught in `Ōlelo Hawai`i~\cite{hoffman2023advancing}. Culturally responsive pedagogy in Hawai`i is primarily based on place-based and āina-based education~\cite{handy1972native, mhasakar2025itrustwesternkumu}\footnote{Āina means ``that which sustains us,'' referring to place as more than land, but also the people, culture, creatures, and environment.}, which ties educational activities to local environments, community knowledge, and Indigenous epistemologies~\cite{harada2016placebased, ngosorio2023ainabased, porter2018alohaaina}. 

In the ethos of community-based participatory research, we worked with public schools in O`ahu with the goal of centering the needs and priorities of educators in the community~\cite{10.1145/2675133.2675147}. Members of our team have years-long relationships with our partner schools, providing a foundation of trust for this project. These relationships shaped key decisions around study design, recruitment, and data collection, ensuring that our process aligned with community values and minimized extractive practices. 

While we focus on public school educators in Hawai`i for this work, our findings speak to broader challenges emerging as generative AI enters classrooms. The specific harms and tool designs surfaced are grounded in this community's values and practices, but the challenges with designing auditing tools in these contexts --- reasoning about cultural misrepresentation in data-scarce settings and designing tools that work across varied forms of expertise (e.g., cultural background, technical / AI literacy) --- are likely to resonate with other Indigenous and marginalized communities. Furthermore, our insights into creating dual-purpose tools that both uncover problematic outputs and serve as pedagogical aids for fostering students’ critical thinking also extend to a wide range of educational contexts. Overall, Hawai`i's cultural and educational landscape makes these dynamics visible in ways that may be harder to detect elsewhere, offering an early window into challenges other communities will soon confront.

\section{Related Work}
\subsection{Algorithmic Auditing}
Algorithm auditing describes the process of surfacing harms, such as social bias, discrimination, or problematic behavior, by investigating outputs of algorithmic systems \cite{birhane2024aiauditingbrokenbus, lacmanovic2025artificial}. The process often involves repeatedly feeding an input to an algorithm, and analyzing its outputs to understand its behavior and subsequent impact \cite{HCI-083}. In recent years, such audit-based methods have proven effective in identifying harmful behaviors when using AI systems in high-stakes domains including housing~\cite{Asplund_Eslami_Sundaram_Sandvig_Karahalios_2020, zou2023ai, juhn2022assessing}, healthcare~\cite{obermeyer2019dissecting, mittermaier2023bias, seyyed2021underdiagnosis}, employment~\cite{chen2018investigating, wilson2024gender, wen2025faire}, and policing~\cite{berk2021artificial, ziosi2024evidence}.    
Although AI audits are often conducted by researchers~\cite{buolamwini2018gender, cramer2018assessing, eslami2019user, eslami2017careful, raji2020closing} or practitioners~\cite{HCI-083, sandvig2014auditing, sweeney2013discrimination}, end users can help identify harms from AI systems by leveraging their lived experiences~\cite{devos2022toward, shen2021everyday}. In fact, in many cases, users can discover issues that expert auditors overlooked, allowing for more comprehensive audits~\cite{mack2024they, mim2024between, shelby2024generative, zhang2024partiality,young2019toward}. Prior work has introduced community-centered auditing practices through focus groups and workshops, demonstrating how this format is beneficial for identifying stereotypical representations of non-Western cultures~\cite{ghosh2024generative} and the lack of disability~\cite{mack2024they} and gender~\cite{ghosh2024don} representation in text-to-image models. \citet{qadri2025casethickevaluationscultural} further illustrated how community-driven evaluation can surface culturally specific and nuanced harms that standard AI evaluation practices miss. Nonetheless, a remaining challenge is how to best support end-users who are engaging in auditing. Currently, practitioners will rely on existing crowdworking platforms for end-user auditing, which may not yield the most productive results~\cite{deng2023understanding}. As an alternative, systems such as \citet{deng2025weauditscaffoldinguserauditors}'s WeAudit and \citet{lam2022end}'s IndieLabel provide examples of tooling built to better scaffold end-users' participation in auditing AI systems for text-to-image generation and toxicity classification respectively. These system artifacts exemplify broader calls to design auditing tools in a more participatory fashion~\cite{ojewale2025towards}.

In parallel, there is growing body of research interested in how we can scaffold end-user auditing processes for target demographic groups. Much of this work has focused on how to support youths in AI auditing, as they are often early adopters of these technologies~\cite{solyst2025investigating,morales2024youth,prabhudesai2025here}. For example, \citet{solyst2025investigating} conducted workshops exploring how to support teenagers in engaging in critical AI auditing work. However, much of this scholarship focuses on feasibility (i.e., demonstrating that youths can productively serve as auditors) rather than on concrete auditing tools or interfaces to best facilitate these processes. Overall, while there is a rich line of prior work that expands individual users' role as auditors, there has been less focus on how to design auditing tools for the specific communities and the harms they experience. Through our co-design workshops, we illustrate how the nuances of the community context in Hawai`i shape the auditing process, from what outputs are considered concerning to what tools are required.

\subsection{Generative AI for Low-Resource Languages and Indigenous Communities}
Prior work has explored how generative AI systems can support the preservation and awareness of Indigenous knowledge systems~\cite{perera2025indigenous}. Efforts include using AI for language revitalization, such as transcribing Indigenous languages~\cite{chaparala2024mai,coto2022development} and promoting language learning~\cite{rahaman2021audio}. The Lauleo project, for instance, crowdsourced audio data in \textit{Ōlelo Hawai`i} (the Hawaiian language) to improve voice-to-text systems~\cite{lauleo2025}. Other work has used generative AI to support cultural education, such as \citet{baker2025kumu}, who developed a system that integrates mo`olelo (Hawaiian cultural stories) and proverbs into lesson plans for Hawaiian language immersion programs.

While AI systems have promising applications for cultural and language education in Indigenous communities, these technologies come with a complex set of challenges. Prior work contends that AI systems are fundamentally rooted in Western traditions and often fail to properly account for Indigenous epistemologies~\cite{lewis2025abundant,maitra2020artificial, bird2024must}. Another challenge arises from data scarcity. AI systems are trained on Western-centric data sources, making them prone to inaccuracies, cultural misrepresentations, and mistranslations for those underrepresented in this data, including Indigenous communities\cite{ahuja2023mega, pinhanez2023balancing, zhong2024opportunitieschallengeslargelanguage}. For Hawai`i, models mistranslate mo`olelo or omit proper `okina punctuation \cite{mhasakar2025itrustwesternkumu}, reflecting a poor grasp of `Ōlelo Hawai`i, a critically endangered language~\cite{pinhanez2023balancing}. These errors can have profound effects on communities. Since Hawaiian culture is passed on via language and stories, which have specific, ancient moral lessons, mistranslations can have generational impacts on understanding of traditions~\cite{malo1903hawaiian}. In addition to concerns about model outputs, generative AI systems raise environmental issues, such as the substantial energy and water requirements required for training and operation~\cite{jegham2025hungry, Bashir2024Climate}. These resource demands compound existing pressures on Indigenous communities~\cite{edwards2025unveiling}. For example, already, there are organizing movements emerging against data centers proposed to be built on Native land~\cite{nodatacenter,boblitt2025data}.

Despite this need for community-led oversight, there are many complexities to scaffolding end-user auditing process with Indigenous communities. Existing auditing approaches often fail to detect harms when auditors lack the lived experiences or cultural knowledge needed to recognize them~\cite{young2019toward}, making Indigenous community involvement critical. Yet many Indigenous researchers remain skeptical of AI’s ability to respect cultural nuance~\cite{birhane2020decolonisingcomputationalsciences}, complicating efforts to conduct user-centered auditing. Without community control, the digitization of Indigenous data can also enable new forms of colonialism, where researchers or companies appropriate cultural resources to train AI without returning benefits to the communities~\cite{ledward2008hcie,young2019new,Spano_Zhang_2025}. Indigenous communities have been vocal about this concern, known as data sovereignty, which outlines how communities should maintain ownership over their data and how it is used \cite{young2019new, cardona-rivera2024indigenous}. Furthermore, prior research has also noted that standard approaches to fairness and bias are often insufficient in Indigenous contexts as they may ignore how a community defines bias~\cite{sloane2022}. Despite these challenges, one recent work has touted the potential of justice-oriented and community-based AI education for achieving data sovereignty \cite{moudalya-2024}. Keeping this potential framework in mind, our work grapples with these complex considerations, providing an example of how to create auditing tools that embed Indigenous knowledge systems and values.

\subsection{Harms of Generative AI in Educational Settings}
Despite the potential boon that deploying generative AI tools have for educational settings, there are corresponding concerns about their potential for harm. In particular, educators are concerned about the broader pedagogical impacts of generative AI systems and the detrimental effects on students' learning, which are further compounded by students' overreliance on AI systems~\cite{harvey2025don, zhai2024effects}. This concern is shared by educators and students alike~\cite{pitts2025studentperspectivesbenefitsrisks}. Prior work has highlighted that overreliance and the subsequent decline in critical thinking skills is especially concerning given AI systems' tendency to hallucinate~\cite{barnum2024we,ji2023survey,gao2022comparing}. This behavior means that educaators must spend extra time to not only learn how to use and integrate these technologies in their classroms but also to check AI outputs for potential errors~\cite{harvey2025don}. Thus, what was initially promised as a time-saver can in fact result in increased labor for educators~\cite{harvey2025don,pangrazio2024data}.

The risks of generative AI systems are exacerbated when deployed with students from historically marginalized communities or in culturally sensitive contexts. These inequalities can manifest in different ways, including the disproportionate focus on the English language~\cite{yan2024practical} or differential performance based on demographic factors such as race or gender\cite{baker2022algorithmic}. Teachers have also expressed concern in AI integration in the classroom exacerbating existing inequalities in educations \cite{harvey2025don}. For example, \citet{gelder-2024}, find that emerging technologies ``don't work'' as intended for Hawaiian immersion teachers unless they are specifically designed with cultural relevance. This misalignment is a harm in it of itself as it undermines educators' agency and devalues local pedagogical curricula. Consequently, while prior work has charted the repercussions of AI tools in educational contexts, less emphasis has been placed on how to best equip educators to discover and address these harms.

\subsection{Equity-Driven Design Methods and Decolonial Computing}
Prior participatory and co-design scholarship has focused on surfacing community priorities, redistributing design power, and mitigating harms when working with marginalized communities~\cite{tseng2025ownershipjusthappytalk, sloane2022, harrington2019deconstructing, cruz2023equityware}. From its origin, participatory design (PD) has supported a democratic approach to addressing social phenomena where power imbalances influence system design~\cite{botero2013ageing}. PD has been applied across diverse contexts, including asset design~\cite{wong2020culture}, civic engagement and community safety~\cite{bailey2017structural, dillahunt2015promise}, collectivist approaches to health inequities~\cite{parker2014collectivistic}, and addressing economic disadvantages~\cite{dillahunt2015promise}. However, participation is not a panacea, and when applied uncritically, PD can reproduce or amplify existing power asymmetries~\cite{harrington2019deconstructing}. To address these concerns, \citet{harrington2019deconstructing} advocate for decolonizing PD practices, emphasizing equity-driven approaches in design.

Related to \citet{harrington2019deconstructing}'s critique of PD, recent HCI scholarship has increasingly turned to decolonial theory to understand and challenge the colonial legacies embedded in computing~\cite{siapera2022ai,mohamed2020decolonial,wong2020decolonizing}. This work moves beyond postcolonial critiques of representation to address \emph{coloniality}, the underlying logic of domination and erasure borne from colonialism that continues to shape the many facets of our lives~\cite{mignolo2007delinking}. 
Within computing, coloniality manifests in the assumption of a universalist, Eurocentric worldview that overlook diverse epistemologies and ways of knowing~\cite{ali2016brief}. 

While the exact definition of ``decolonial'' varies across works in HCI and critical computing, these works share a unifying ethos focused on embracing pluriversality, or sustaining a ``world of many worlds,'' rather than designing for a single universal standard~\cite{lazem2022challenges,wong2020decolonizing}. Applying decolonial thinking to computing systems has led to critiques of technologies that enforce these universalist perspectives, such as content moderation on social media platforms~\cite{shahid2023decolonizing}, digital mental health practices~\cite{pendse2022treatment}, or data collection in natural language processing (NLP)~\cite{held2023material,bird2020decolonising}. Other works highlight how Indigenous and Global South communities repurpose technologies, originally designed with Western defaults, to better fit local knowledge and values~\cite{millan2024-cosmovision,kotut2022winds,bidwell2021decolonising}. Beyond providing an analytical lens, decolonial theory has spurred researchers to consider what these systems would look like if they were designed in a fundamentally decolonial manner, such as those that emphasize the value of ``care''~\cite{shahid2023decolonizing,villalobos2025interaction}. We build on this scholarship to inform recommendations around how we can design end-user auditing processes that support the plurality of needs, ways of knowing, and values across different communities.

\section{Methods}
\subsection{Participants and Sites} 
Our co-design workshops were hosted in-person at two public elementary schools in O`ahu, Hawai`i during March 2025. Of the two schools we hosted workshops at, one offers a Kaiapuni (Hawaiian language immersion) program. We recruited participants through partnerships with a local education organization, comprising of educators and cultural experts based in Hawai`i. Through our partners, we circulated advertisements to educators via internal mailing lists at each of the partner schools. The demographics of workshop participants are reported in Fig.~\ref{fig:uses}; to preserve the anonymity of participants we provide only aggregate statistics. Workshop participants were compensated with a \$40 gift card. This project was approved by Stanford University's Institutional Review Board.

\subsection{Workshop Activities}
In total, we ran four one-hour workshops with 22 participants facilitated by two members of the research team. Each session hosted a different set of participants: five in Workshop 1, four in Workshop 2, seven in Workshop 3, and six in Workshop 4 (22 in total). To establish a shared understanding and vocabulary of generative AI's applications and biases across participants, all workshops started with a short presentation from the facilitators; this presentation was identical across all workshops. After, participants engaged in a design exercise. Participants in Workshops 1 and 2 completed a rapid prototyping exercise, whereas in Workshops 3 and 4, participants used storyboarding. The workshops used the following agenda:
\begin{enumerate}
    \item After introduction and setting workshop norms, participants were invited to share how they are currently using generative AI in the classroom (if at all). The researchers on the team then provided a \textbf{brief presentation about different educational applications} of generative AI, covering both teacher-facing (e.g., making lesson plans, creating rubrics) and student-facing (e.g., creating interactive activities) use cases, since not all participants had prior experience with these technologies.
    \item Second, we invited participants to \textbf{reflect on concerns} they had with using generative AI in educational settings. We asked open-ended questions about their experiences, including whether they had personally encountered or heard of instances of misrepresentation, bias, or other harms. Importantly, participants were free to raise any topics they considered relevant, and the range of concerns extended beyond issues we introduced. 
    \item In the third part, participants were shown three \textbf{design probes} and were asked to engage in open discussion to share their thoughts. The participants were not explicitly prompted to focus on any types of harm, including cultural representation.
    \item Finally, we gave a brief presentation on AI auditing with examples about what purpose auditing serves and what types of inputs / outputs are expected from auditing tools. Participants then engaged in one of two \textbf{design exercises} to prototype auditing tools they would find useful for identifying problematic outputs when using generative AI. All participants shared their designs with the group, fostering further discussion on the types of auditing tools participants wanted.  
\end{enumerate}
At the conclusion of the workshop, participants were given a brief survey, covering their concerns with using generative AI tools, reflections from the workshop, and demographic information. 
\begin{figure*}[t!]
    \centering
    \includegraphics[width=\textwidth]{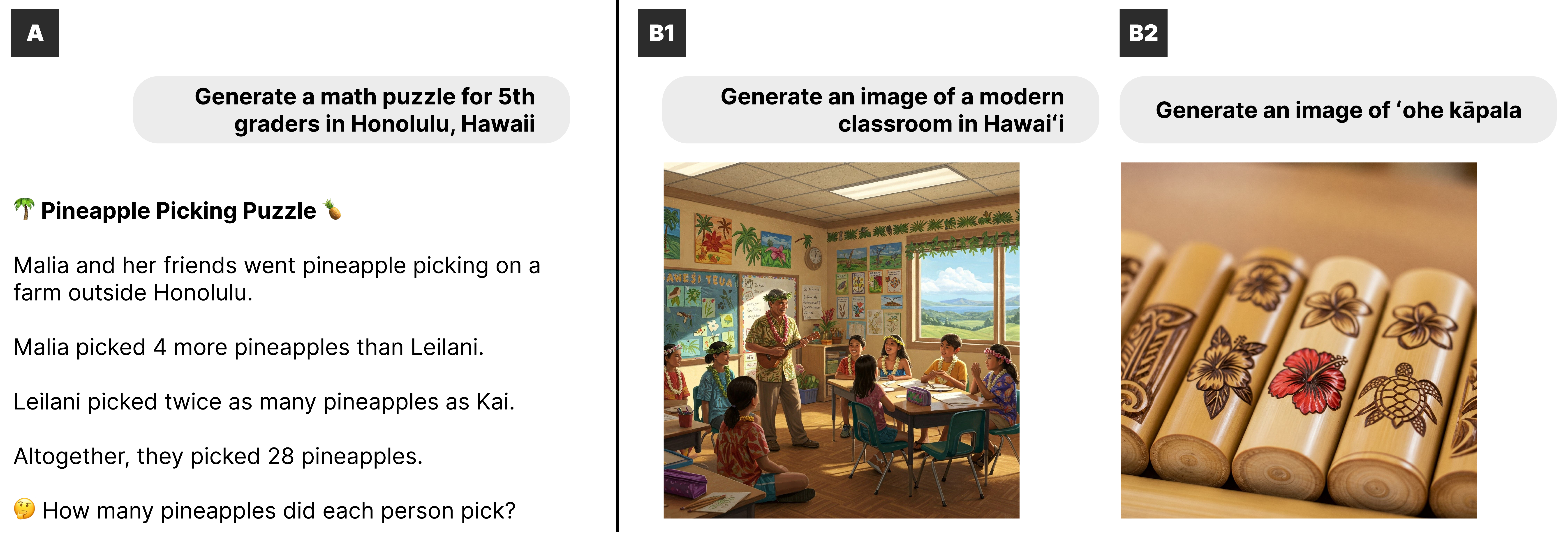}
    \caption{Examples of design probes shown to participants during the design workshop. In addition to the example with Captain Cook shown in Fig.~\ref{fig:teaser}, participants were shown probes related to creating assessment questions (left) and generating images (right). These outputs show actual outputs from ChatGPT and Gemini from March 2025. For the image generation probe, we asked the model to provide two images: the first was of a ``modern classroom in Hawai`i'' (2A) and the second of an `ohe kāpala, a flat bamboo stamp typically carved with geometric patterns used to decorate clothing (2B).}
    \label{fig:probes}
    \Description[Examples of design probes presented to participants during the workshop.]{Examples of design probes presented to participants during the workshop. Panel A shows a math word problem generated by ChatGPT for 5th graders in Honolulu, framed as a ``Pineapple Picking Puzzle." Panel B shows two image generation outputs: B1 depicts a ``modern classroom in Hawai`i,'' illustrated with students seated around tables, tropical posters on the walls, and a teacher playing the ukulele; B2 depicts a round wooden stamp with an elaborate flower pattern.}
\end{figure*}
\subsubsection{Design Probes} 
For each workshop, we presented participants with the following probes, which are intentionally designed to match how generative AI is commonly used in educational settings~\cite{gallup2025teaching}. Our probes were developed across two iterative feedback sessions with educators in Hawai`i and depict actual outputs generated from ChatGPT and Gemini in March 2025. The three probes are as follows:
\begin{itemize}
    \item \textbf{Localizing quiz generation}: Our first probe presents an example of an educator creating a math problem for fifth grade students based in Honolulu, Hawai`i, a common educational use case for generative AI system (Fig.~\ref{fig:probes} A). The generated quiz question setup is around ``pineapple picking,'' presenting a stereotypical and offensive example given the role of pineapple plantations in Hawai`i's colonization.
    \item \textbf{Chatting with a historical figure}: The second probe presents a scenario where a student is interacting with a generative AI system simulating a historical figure, a feature offered by existing educational AI platforms.\footnote{https://app.schoolai.com/spaces/clmqu2ycm00g93b664ot10jq9} The student converses with Captain James Cook, who presents his exploits in Hawai`i as feats of navigation as opposed to acts of colonization (Fig.~\ref{fig:teaser}).
    \item \textbf{Generating images}: Finally, our third probe covers text-to-image generation (Fig.~\ref{fig:probes}), which educators may use when creating instructional materials. We generated two initial images: the first image depicting a ``modern classroom in Hawai`i'' (Fig.~\ref{fig:probes} B1) and the second an `ohe kapala, a flat bamboo stamp used in Hawai`i for fabric-making that students learn to craft themselves as part of arts courses (Fig.~\ref{fig:probes} B2). Participants were allowed to explore additional images. In Workshop 2, participants generated an image depicting \emph{hukilau}, a traditional Hawaiian fishing method. 
\end{itemize}

\subsubsection{Design Exercises} 
We conducted two exercises intended to facilitate divergent and convergent design thinking for participants. In the first two workshops, participants engaged in a rapid prototyping exercise intended to generate divergent designs~\cite{knapp2016sprint}. Participants were given eight to ten minutes and asked to ideate as many ideas as possible for tools that would support them in discovering problematic outputs of generative AI systems. At the end of the time window, individuals presented their ideas to the workshop to seed further discussion on tool designs. In our second set of workshops, our goal was to refine upon these set of tool ideas. Participants formed groups of two to three and selected one of the previously generated tool ideas for a storyboarding exercise. For the selected tool idea, participants were asked to sketch panels detailing what the educator is using AI for, what types of cultural misrepresentation the teacher is concerned about, drawing an example of an AI auditing tool they would use, and listing what the auditing tool reveals about the AI system that would make them comfortable to use it (or decide to not use it). Similar to the first set of workshops, at the conclusion of the exercise, each group shared their storyboard to facilitate further reflective discussion.

\subsection{Analysis}
\label{subsec:analysis}
We audio-recorded and transcribed three of the four workshops. For the remaining workshop (Workshop 3), researchers took detailed notes, which were included in the analysis. We adopted an iterative, inductive approach to thematic analysis with three members of the research team iteratively coding workshop transcripts and produced design artifacts (i.e., rapid prototyping sketches and storyboards). We first independently generated initial codes capturing salient ideas, which were organized on Miro, a collaborative whiteboarding tool. These codes were then grouped into broader themes over two synchronous meetings; these themes formed the basis of our codebook (see Appendix~\ref{sec:app_codebook}). The same three researchers applied this codebook independently analyzing all workshop transcripts and design artifacts. We met synchronously after coding each transcript to refine theme definitions, add missing codes, and negotiate differences in interpretation. Each workshop transcript is coded independently by all three researchers. After analyzing all transcripts, we synthesized the exemplar quotes from the process, removing any duplicates that emerged during coding. 

We adopted a reflexive thematic analysis approach to analyze the transcripts as our goal was to explore themes grounded in participants’ experiences, which was important given the emergent and relatively underexplored nature of the workshop topics~\cite{clarke2014thematic}. Following existing practices for inductive approaches, we do not report inter-rater reliability as the goal of our analysis is surfacing salient patterns from the workshops rather than quantifying how often these themes arose~\cite{mcdonald2019reliability}. Instead, the three authors analyzing the transcripts established a shared understanding of the meaning and application of themes through rounds of discussion and reflection. To preserve anonymity, we identify participants only by workshop number and omit personal identifiers.

\subsection{Positionality}
Our authors comprise a team of nine people with a range of backgrounds germane to cultural representation in Hawai`i and AI auditing practices. The team consists of members with experience on algorithmic approaches to measuring and mitigating social biases in ML systems, community-engaged design, and AI for education. Two of our authors are of Kanaka Maoli descent; one of the authors identies as Kama`āina (local) and is a former public school educator in Hawai`i; the remaining authors are cultural outsiders. Our workshop materials were reviewed by all members of the research team. Several team members have existing collaborations with the partner schools in O`ahu, helping coordinate our workshops. The workshop materials were co-developed by all members of the research team, and coordination with the research sites was led by senior authors on the team who had established relationship with schools in Hawai`i. The first and third authors (both cultural outsiders) conducted the workshops; in addition to the first and third authors, the second author (also a cultural outsider) developed the codebook and analyzed the transcripts. Finally, all members of the research team were invited to contribute feedback on the extracted themes and resulting draft.

\section{Findings}
In this section, we present learnings from our workshops. First, we situate how participants are engaging with AI systems, exploring how they are currently using or want to use generative AI in educational settings. Next, we discuss the categories of harm --- focusing on dimensions of cultural representation --- that teachers are most concerned about. This understanding helps us localize what types of harms participants are likely auditing for and in what scenarios these tools will likely be used. We next discuss design features participants proposed for auditing tools, mapping them to harms. Finally, we provide a design vignette of an exemplar system created as a result of the co-design sessions.

\subsection{How are teachers engaging with generative AI systems?}
\label{subsec:usage}
\begin{figure*}
    \centering
    \includegraphics[width=\linewidth]{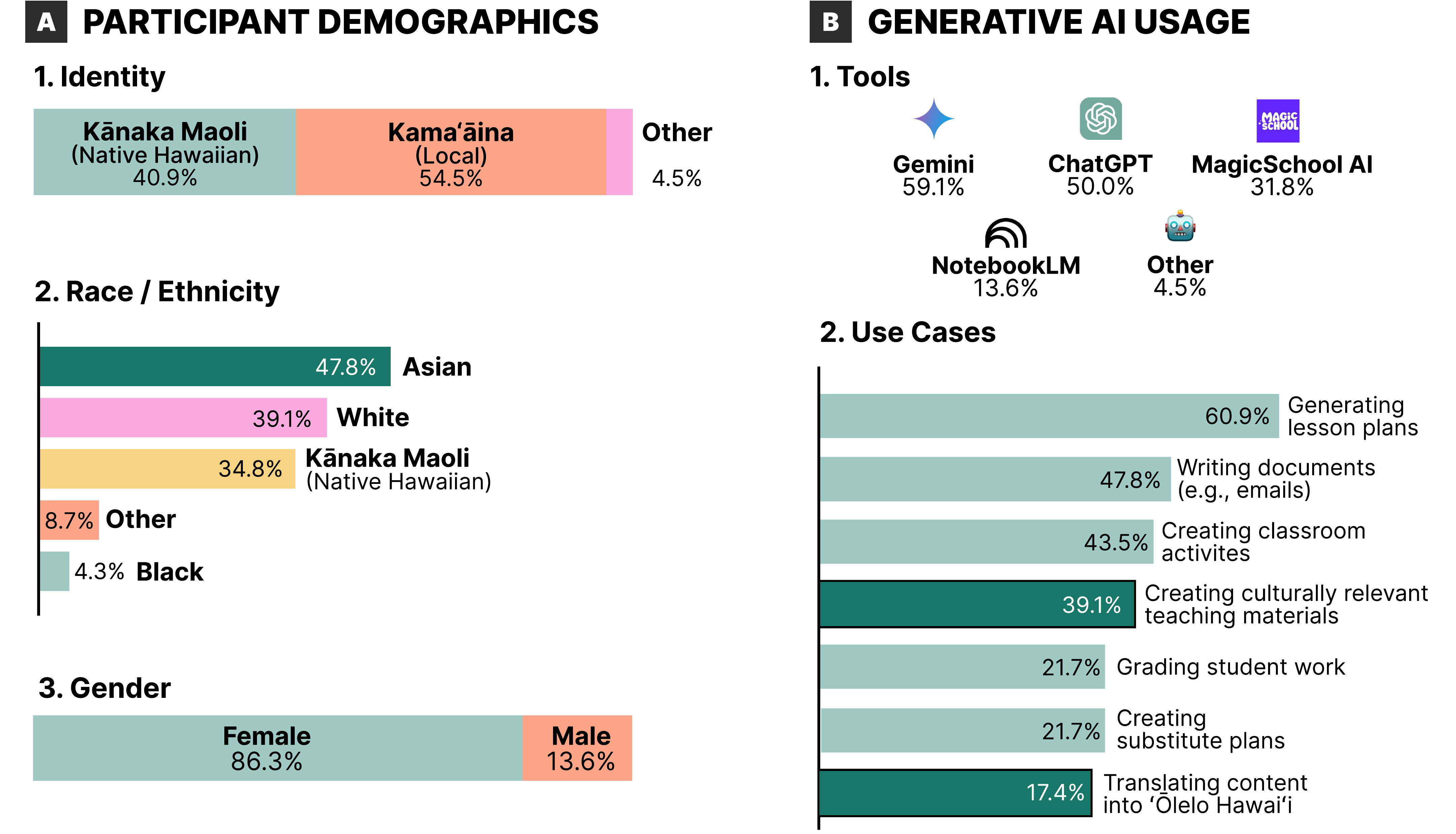}
    \caption{We report summary statistics on our 22 workshop participants, including information on their demographics (A) and generative AI usage (B). Of our 22 participants, all but one had used AI tools in the past and 68.1\% (N=15) had specifically used AI tools in educational settings. We find that educators had most frequently used AI for logistical tasks such as generating lesson plans and writing documents. While usage patterns in our sample broadly mirror those from surveys conducted with K-12 educators across the U.S.~\cite{gallup2025teaching}, approximately $40\%$ of participants reported using AI tools for tasks related to Hawaiian culture (as noted in the dark green outlined bars in B2), which includes generating lesson materials and translating content.}
    \label{fig:uses}
    \Description[ Participant demographics and generative AI usage.]{
    Participant demographics and generative AI usage. Panel A shows demographics of the 22 public school educators in the study. By identity, 40.9\% identified as Kānaka Maoli (Native Hawaiian), 54.5\% as kama‘āina (local), and 4.5\% as other. By race/ethnicity, 47.8\% identified as Asian, 39.1\% as White, 34.8\% as Kānaka Maoli, 8.7\% as other, and 4.3\% as Black (participants could select multiple categories). By gender, 86.3\% identified as female and 13.6\% as male. Panel B shows educators’ generative AI usage. The most commonly used tools were Gemini (59.1\%), ChatGPT (50.0\%), and MagicSchool AI (31.8\%). Reported use cases included generating lesson plans (60.9\%), writing documents such as emails (47.8\%), creating classroom activities (43.5\%), creating culturally relevant teaching materials (39.1\%), grading student work (21.7\%), creating substitute plans (21.7\%), and translating content into ʻŌlelo Hawai‘i (17.4\%).
    }
\end{figure*}
First, we sought to understand how participants use generative AI in their day-to-day roles as educators. Of the 22 participants, all but one reported having used AI in the past, and $68.2\%$ (N=15) specifically used AI in their teaching practices. As shown in Fig.~\ref{fig:uses} B1, the most commonly used generative AI tools among teachers included general-purpose tools, such as Gemini (N=13) and ChatGPT (N=11), as well as education-specific tools like MagicSchool AI (N=7). Many participants reported using AI systems to create lesson plans (N=14), generate impromptu classroom activities (N=10), or prepare materials for stand-in teachers (N=5). Some also used AI for tasks requiring knowledge of Hawaiian culture and practices, such as developing culturally relevant teaching materials or translating content into \textit{`Ōlelo Hawai`i} (Fig.~\ref{fig:uses} B2). Overall, our participants' usage patterns mirror results from recent work surveying how K-12 teachers in the U.S. more broadly are employing generative AI tools~\cite{gallup2025teaching}. However, using these tools to assist developing culturally relevant educational material is unique to the sample we observe.

We asked participants whether they had encountered any cases where AI-generated outputs did not align with the history or perspectives they been taught or want to teach. Half of our participants reported personally encountering instances of bias in the past when using generative AI tools for educational purposes. For example, a participant in Workshop 1 found the model's knowledge of Hawai`i to be surface-level: ``\emph{[generated outputs] are very mainstream...it knows the ones that most other people know like Aloha Oe but beyond that, when you're looking for kind of more specific things it's a little bit lacking in that area.}'' Another participant in Workshop 3 mentioned that when they tried to generate images related to Hawai`i, the output always contained people in hula skirts even when not specified. Finally, another participant in Workshop 4 mentioned how generated outputs have contained \textit{``incorrect facts about Hawaiian history.''}

\subsection{What concerns should AI auditing tools address?}
\label{sec:concern}
\begin{figure*}[ht!]
  \centering
\includegraphics[width=\textwidth]{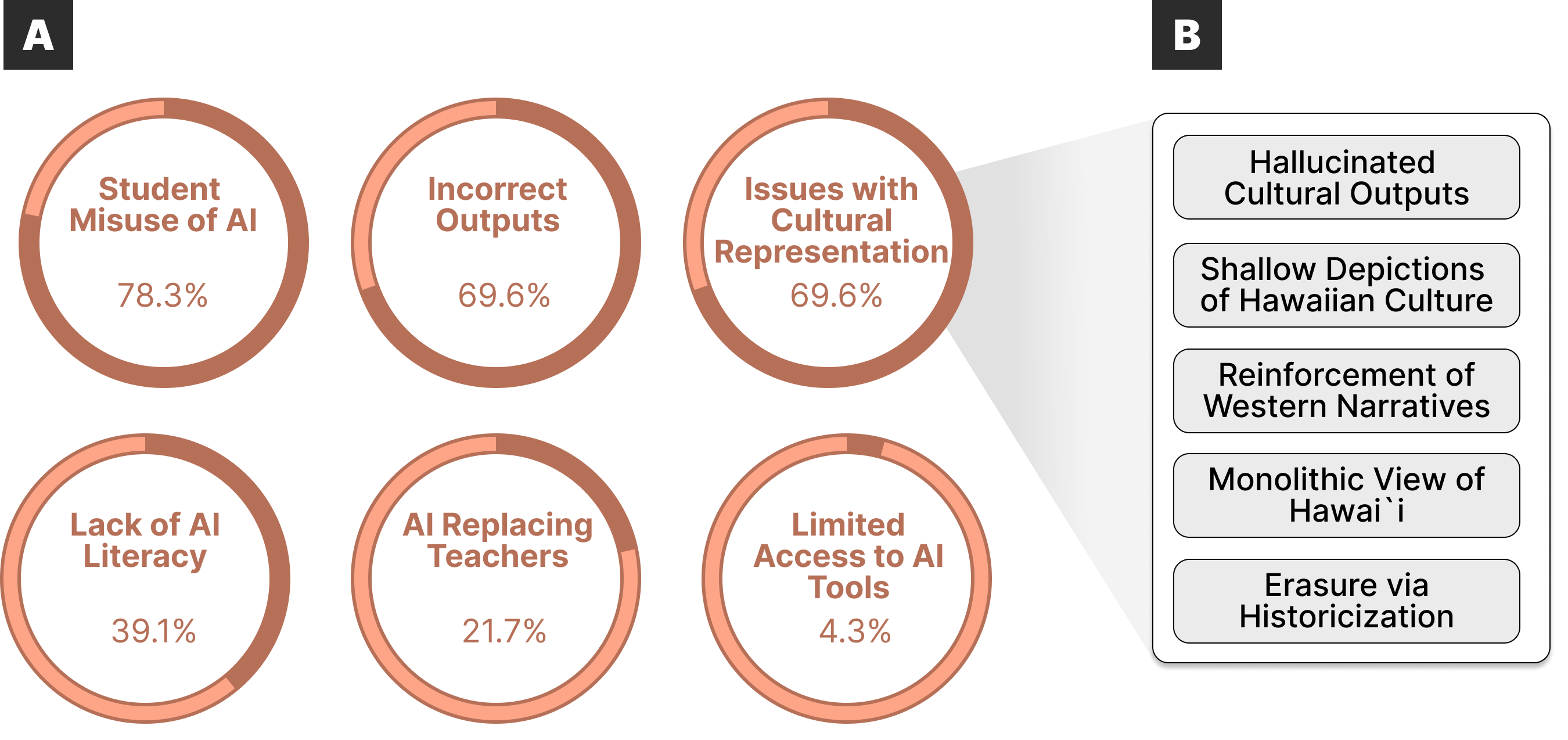}
  \caption{Educators are concerned that AI systems will misrepresent aspects of Hawaiian culture. We report survey results on what concerns workshop participants have with using AI in educational contexts (A). From our thematic analysis, we provide five dimensions of cultural misrepresentation that our participants believe AI auditing tools ought to address (B).}
  \label{fig:ai_tool_concerns}
  \Description[Panel A presents survey results on participants’ concerns. Panel B shows five dimensions of cultural misrepresentation identified through thematic analysis.]{Panel A presents survey results on participants’ concerns. The most frequent issues were student misuse of AI (78.3\%), incorrect outputs (69.6\%), and issues with cultural representation (69.6\%). Other concerns included lack of AI literacy (39.1\%), AI replacing teachers (21.7\%), and limited access to AI tools (4.3\%). Panel B shows five dimensions of cultural misrepresentation identified through thematic analysis: hallucinated cultural outputs, shallow depictions of Hawaiian culture, reinforcement of Western narratives, monolithic views of Hawai‘i, and erasure through historicization.}
\end{figure*}

Next, we examined the types of concerns participants raised about using generative AI and what harms audits should address. While participants surfaced a wide range of concerns --- from student misuse of AI (e.g., academic dishonesty, inhibited learning) to broader societal impacts (e.g., job loss) --- we focus on issues with cultural representation in education settings, where auditing tools are especially well-suited to intervene. We identified five recurring concerns: (1) inaccurate outputs, (2) superficial cultural representation, (3) dominance of Western narratives, (4) lack of diversity, and (5) framing Hawaiian culture as historical rather than living. For each dimension, we discuss the pedagogical consequences participants linked to each concern. Finally, we report how participants’ broader concerns about generative AI in education intersect with those related to cultural representation.

\subsubsection{Presenting incorrect or hallucinated outputs}
The most common concern that participants expressed pertained to hallucinated outputs, especially related to culturally incorrect information. This finding is consistent with prior works that have found users are most familiar with hallucinations as a form of harm or misrepresentation from generative AI systems~\cite{qadri2025casethickevaluationscultural}. In our case, participants were especially concerned about the tangible repercussions hallucinated outputs could have on their students’ education. For example, in Workshop 1, a participant mentioned how generated outputs would conflate parts of different mo`olelo, or Hawaiian cultural stories: 
\begin{quote}
     I used ChatGPT to organize the different mo`olelos throughout the year... Like there was one about Maui and how we read about how to find the Hawaiian islands. And then I noticed in ChatGPT --- wait a second --- this is talking about Maui when he gets the sun. This isn’t the same mo`olelo that they read.
\end{quote}
This participant was particularly concerned as learning mo`olelo was a core aspect of state-level standards for their students.

Participants also noted inaccuracies with language generation. For example, participants found that AI-generated phrases in `Ōlelo felt inconsistent with how native speakers write (``\textit{certain phrases you would never really use in a letter, but [the AI tool] uses that}'' [W2]) or are missing `okina, a symbol commonly used in `Ōlelo. These inaccuracies pose a particular challenge for teachers, such as those in Workshop 4, that are part of the Kaiapuni program (Hawaiian language immersion) in which classes are taught exclusively in `Ōlelo. 

\subsubsection{Providing a surface-level depiction of Hawaiian culture}
A second dimension of cultural misrepresentation that participants discussed was related to the specificity of representation that occurred in both existing educational materials and generated outputs~\cite{qadri2025casethickevaluationscultural}. Educators mention a pedagogical concern that existing educational materials tend to cater towards the cultural experiences of those on the mainland, presenting concepts that were irrelevant or unfamiliar to students. For instance, participants gave examples of provided educational materials or standardized testing frequently making mentioned to concepts, such as ``\emph{snow days} (W1)'', that their students had not heard of before: ``\emph{or the attic, [standardized tests] always talk about playing in the basement and my kids are always like, what’s a basement? We don’t have those.} (W1)'' Participants emphasized that these irrelevant examples would appear ``\emph{in the middle of high-stakes testing}'' (W1), and the added confusion for students could have negative repercussions on their academic performance~\cite{sundararajan2020keep, park2015cognitive}. This disconnect highlights a potential opportunity for generative AI tools, which could ideally be used to generate personalized content that resonates with students' local experiences.

While generative AI presents the opportunity to localize education materials, participants note that the resulting outputs are superficial, reproducing the pedagogical repercussions they aim to avoid. Participants reported that AI outputs often provided a ``mainland'' perspective of Hawaiian culture that was incongruous with students' lived experiences and could even perpetuate harmful stereotypes. For example in Workshop 3, participants found the design probe describing a math problem containing pineapples to provide a shallow reference to Hawaiian culture; they mentioned that they frequently encountered generated AI outputs with pineapple even though ``\emph{[pineapples] don't represent us at all}'' (W3). Furthemore, participants pointed out that pineapples could be considered offensive given the fruit's association with the colonization of Hawai`i~\cite{kleinkingdom}. As an alternative, participants wanted the generated output to use artifacts germane to their students' lives, such as \emph{kalo} (taro). Thus, while generative AI tools hold promise for adapting learning content for Hawaiian students, this personalization risks becoming another form of misrepresentation and enacting the same pedagogical consequences.

\subsubsection{Reinforcing Western narratives}
It is not only what is in the outputs but whose narratives and viewpoints are showcased. Participants discussed how generative AI tools presented a Western perspective of history, glossing over historical injustices or harm to Hawaiian communities. For example, a participant in Workshop 1 highlighted how the outputs of generative AI tools skew in favor of Western colonialism of Hawai`i, even going as far as omitting crucial Indigenous history entirely: ``\textit{you’re only going to read it from the perspective of the [U.S.] government, right? They came in to save the day. But where's the perspective of Queen Liliuokalani?}'' A participant in Workshop 3 made the same point, stating ``\textit{we know certain things happened but AI said it hadn't}.'' Predominantly Western-centric curricula and materials can not only marginalize Indigenous perspectives, but lead to the underachievement and disengagement of Indigenous students~\cite{vallee2018eurocentrism, riley2024weaving}. These erasures of Hawaiian history actively undermine students' opportunities to engage critically with history, presenting a significant barrier to culturally grounded learning~\cite{fricker2007epistemic}. 

\subsubsection{Failing to showcase diversity}
One subset of stereotyping that participants were particularly concerned about related to outputs' inability to capture the true diversity of Hawai`i. As mentioned in Sec.~\ref{sec:background}, Hawai`i has a large multicultural population, yet as participants in Workshop 2 noticed when observing generated images of classrooms in Hawaii`, ``\emph{these pictures all seem to just be like Polynesians, where Hawai`i is so much more than just Polynesians.}'' Similarly, a participant from Workshop 1 commented that ``\emph{it's interesting that there's not a white child... like the assumption there that everybody here is of Hawaiian descent or even a mix, but look at our school population --- many students are white, right? Like 30\% or something}.'' The overrepresentation of Hawaiian individuals in generated outputs can help prevent cultural erasure~\cite{qadri2025casethickevaluationscultural}. However, it risks flattening the cultural diversity in Hawai`i that more closely aligns with students and teachers' lived experiences. 

\subsubsection{Representing Hawaiian culture as historical rather than living} 
Participants noted that AI outputs often portrayed Hawaiian culture as belonging to the past rather than the present. In response to a design probe showing an AI-generated classroom in Hawai`i, participants described it as ``\emph{old-fashioned}'' and ``\emph{like how it would look in the 80s}'' (W2). Similarly, the initial image of a \emph{hukilau}, a Hawaiian fishing practice, that participants in Workshop 2 generated was rendered first as a stylized historical line drawing; only after iterating on the prompt were participants able to generate more photorealistic and contemporary depictions. These observations reflect broader concerns about how K-12 education often engages in this \emph{whitewashing}, where Indigenous people and history are framed as passive or ``\emph{relics of the distance past}''~\cite{shear2015manifesting}, rather than existing in the modern context.

\subsubsection{Intersecting concerns around generative AI use}
Finally, participants’ concerns were not isolated; issues of cultural representation often intersected with broader worries about \remove{how} generative AI usage in educational settings. Many educators focused on how cultural misrepresentations directly affected their students, which amplified existing concerns about student misuse of AI. While participants felt confident in their own ability to identify harmful representations, they were concerned that students lacked the same critical awareness to evaluate generated outputs. As one participant explained, ``\textit{[students] are just believing whatever they see, you know?... It's hard because you just believe it, right? Whatever you read, you believe. And a lot of it is biased}'' (W1). In addition, participants were concerned that students would become overreliant on generative AI. For example, a participant in Workshop 3 stated that they try to teach students that generative AI is ``\emph{a resource but not the only resource}.'' For students learning about Hawaiian culture, they emphasized the importance of seeking out other methods, such as speaking to community elders or learning stories that are traditionally passed down orally.

Outside of students interacting with generative AI systems, participants also highlighted the compounding effects of limited AI literacy. For example, when encountering hallucinated or stereotypical outputs, educators wondered if the problem stemmed from how they prompted the model. Some participants stated that they did not even know how to prompt models to get culturally relevant information. Furthermore, concerns about cultural misrepresentation often left participants feeling they had to ``\emph{double-check}'' (W4) every AI-generated response. Ironically, these technologies marketed as time-saving may increase educators' workloads instead of reducing them~\cite{harvey2025don,selwyn2025prompting}.

\subsection{What functionalities do users want in AI auditing tools?}
\label{subsec:wireframes}
\begin{figure*}
    \centering
    \includegraphics[width=\textwidth]{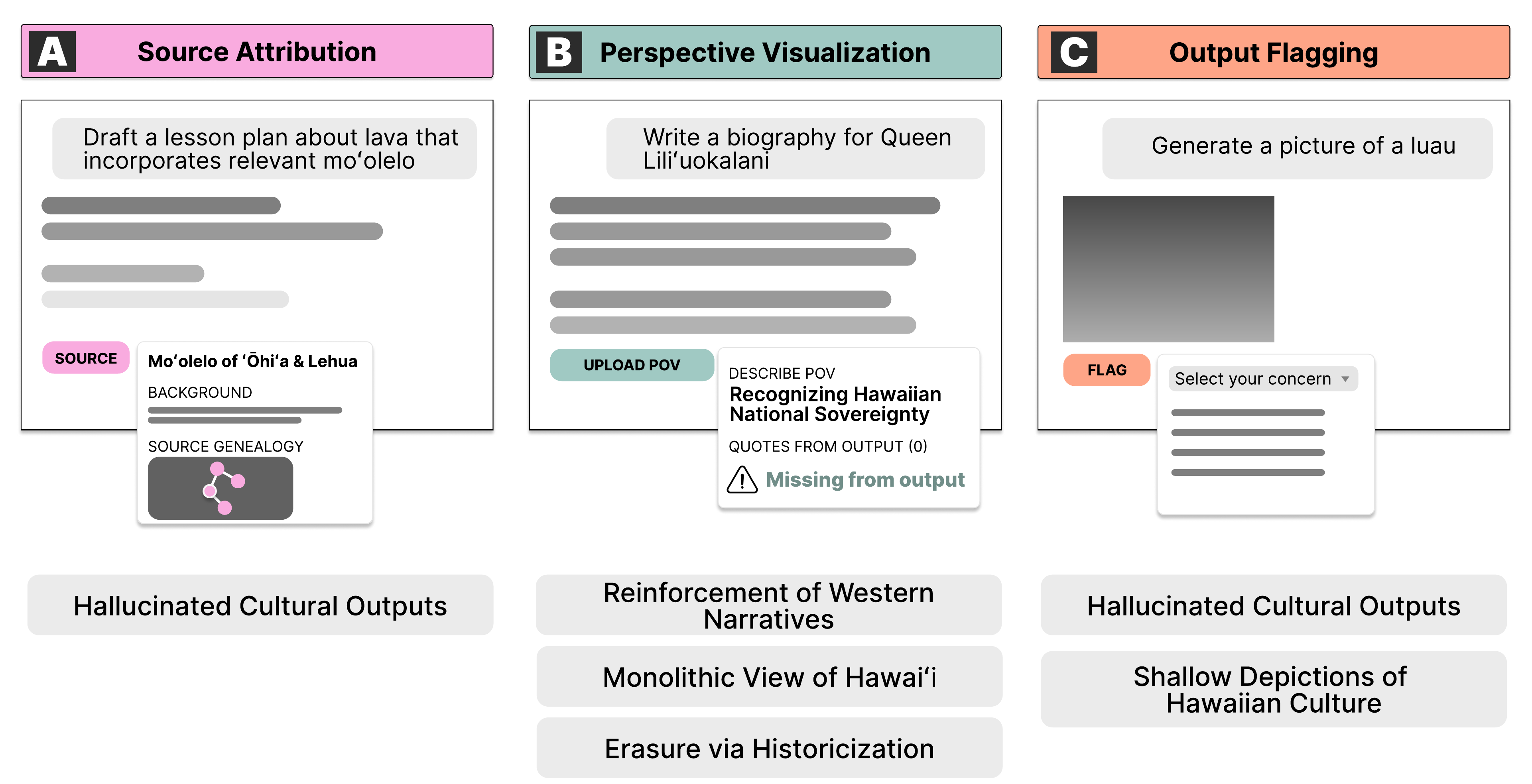}
    \caption{We present wireframes of auditing features based on the sketches and storyboards that participants created during the co-design sessions. The three design elements that occurred consistently across co-design workshops included source attribution (A), perspective visualization (B), and flagging problematic outputs (C). We map each wireframe to the participant concerns identified in Sec.~\ref{sec:concern} that the feature was designed to address.}
    \label{fig:wireframes}
    \Description[Wireframes of auditing features created from participants’ sketches and storyboards.]{
    Wireframes of auditing features created from participants’ sketches and storyboards. Panel A illustrates source attribution, where a lesson plan output can be traced back to relevant mo‘olelo (stories), with a genealogy view showing background and connections. Panel B illustrates perspective visualization, allowing users to upload alternative points of view—such as recognizing Hawaiian national sovereignty—to highlight perspectives missing from an AI-generated biography. Panel C illustrates output flagging, where users can flag problematic outputs, such as a generated luau image, and specify the nature of the concern. Each wireframe is mapped to participant concerns identified earlier: hallucinated cultural outputs, shallow cultural depictions, reinforcement of Western narratives, monolithic views of Hawai‘i, and erasure through historicization.
    }
\end{figure*}
Finally, we present three categories of auditing tools ideated by participants during our workshops. As illustrated in Fig.~\ref{fig:wireframes}, these include: (1) tracing sources, (2) identifying perspectives in generated outputs, and (3) flagging instances of harmful content. While these tools resemble existing, general-purpose approaches for auditing or mitigating hallucinations in generative AI, our findings show that participants’ designs are shaped by the cultural context and educational setting. For each design, we describe the underlying concerns it addresses as well as key considerations for implementation.

\subsubsection{Identifying sources}
\label{subsec:sources}
As shown in Fig.~\ref{fig:wireframes}A, participants in Workshop 4 proposed an auditing tool that allows users to trace the genealogy of a linked source, displaying information about the author as well as the lineage of their teachers. Participants stressed that understanding where outputs are drawn from was essential for verifying whether they could trust outputs. But here, source verification is not simply attributing a citation to the output. As participants explained, in the context of cultural knowledge in Hawai`i, there is often no definitive ground-truth to compare outputs against since cultural artifacts, such as mo`olelo or proverbs, will differ depending on the practitioner's interpretation. As a participant in Workshop 4 stated:
\begin{quote}
    ``\emph{Someone can take a proverb and write a whole different aspect, so it is just about who's your teacher? ... You type in [the author's] name, who was their kumu? You're always going back to who was their teacher? Because other than that, you wouldn't trust it.}''
\end{quote}
Because of these interpretive differences, participants described that it was often more valuable to understand the \emph{genealogy} of knowledge --- who authored a source and who taught them --- than to evaluate the content of a generated output directly. Unlike traditional fact-checking or citation methods, which presuppose there is a correct or incorrect answer, this design enables the more interpretive work that participants described as essential for assessing culturally grounded knowledge. 
 
How to implement such a feature presents challenges regarding technical feasibility. Knowledge attribution for generative AI is an active area of research within NLP~\cite{khoo2022deepfake, wang2024sourceattributionlargelanguage, tilwani2024neurosymbolicaiapproachattribution}. Our setting provides additional challenges as conventional citation based systems often rely on digitized and unambiguous references, whereas sources here may not be digitally accessible or interpretative. Furthermore, this tool would require creating and maintaining a corpus of culturally validated material, as well as require users to contribute to its source genealogy, would require significant community labor and governance structures for who is allowed to contribute validated sources.

Another critical consideration is regarding who has control and access to the curated data. This consideration is tied to the rich discussion on Indigenous data sovereignty surfaced in the existing literature, particularly in the context of AI systems~\cite{birhane2020decolonisingcomputationalsciences, cardona-rivera2024indigenous}. We see similar concerns about the misappropriation of cultural resources in our workshops as well. For example, in Workshop 4, participants mentioned they fear sharing resources they have created for the Kaiapuni program on the Internet out of fear it might be criticized or misused. For instance, they mentioned online figures who are ``\emph{totally anti-Hawaiian and flip everything around and use Hawaiian culture against us}'', or examples of newspapers that publish in the Hawaiian language but are incongruous with Hawaiian culture such as claiming that \textit{```ohana is a made up word''} (W4). Thus, for this design to be successful, it most not only integrate with Hawaiian epistemologies that center knowledge provenance, but also uphold data sovereignty, ensuring any source material or cultural interpretations collected remains in the hands of the community.

\subsubsection{Visualizing perspectives}
Stemming from concerns that models reinforce Western narratives and provide a narrow depiction of Hawai`i, participants in Workshops 1 and 2 expressed interest in visualizing different viewpoints within AI-generated outputs (Fig.~\ref{fig:wireframes}B). Participants wanted outputs that could showcase a multiplicity of perspectives. One proposed design was a lightweight visualization that displays the distribution of perspectives in a response: ``\emph{How do you know that you’re getting [a generated output] that is well-rounded... I want like a little visual that's like `this [output] is solid'}'' (W1). Equally important, participants stressed, is visibility into omissions: ``\emph{it would be interesting to be able to identify what perspectives went into that [generated] biography and what perspectives are missing}'' (W2). Overall, these auditing tools are especially critical given that existing curricula and teaching materials already tend to foreground mainland perspectives (see Sec.~\ref{sec:concern}). Educators already have to expend extra effort to adapt these materials for their classroom~\cite{yong2014teacher}. If AI systems are intended to help localize educational content~\cite{baker2025kumu,grab2025teaching}, it becomes crucial to ensure they do not reproduce the same biases and compound existing inequities. 

Achieving such a visualization requires several technical considerations. Perspective auditing is closely related to the classic NLP task of stance classification \cite{kuccuk2020stance}, which infers whether text supports or opposes a given claim. One approach to perspective visualization is combining topic modeling~\cite{deerwester1990indexing}, to surface the key issues in a text, with stance classification, to detect alignments on those issues. Detected stances can then be compared against a repository of known perspectives, either curated directly or gathered from external sources. These approaches are helpful for visualizing perspectives include in text, but understanding what perspectives are missing is an infinitely large design space. A tractable approach is to have users define a perspective and then check if it is present in an output, although this method assumes the user knows what perspectives they believe ought to be included.

One important distinction to make with such a feature is distinguishing when to showcase a plurality of perspectives versus actively addressing historical erasure~\cite{Mollema_2025}. Framing erasures or false Western narratives as mere ``perspective'' differences can undercut the severity of the issue. If we fail to acknowledge why Hawaiian perspectives have been historically suppressed, these auditing features can actually undermine efforts to address epistemic injustices~\cite{tallbear2019caretaking,fricker2007epistemic}. Even the notion of what constitutes a perspective demands careful consideration: if defined too broadly, it can flatten the rich diversity within a community and treat distinct voices as a monolith. For example, Workshop 4 participants noted that even Hawaiian-language newspapers offered contrasting editorial stances depending on who was in charge. As one participant explained, ``\emph{Sometimes when you look at the Hawaiian newspaper and you read it, you're like, what am I reading? You know, it's all in Hawaiian, but they're telling you, oh, `surfing is bad'.}''

\subsubsection{Flagging problematic outputs}
Finally, participants in Workshop 1 suggested tools for flagging instances of cultural misrepresentations (e.g., hallucinations, stereotyping) and other problematic outputs. Whereas the previous designs were centered around helping participants identify issues with generated outputs, participants were interested in using this feature for raising awareness. In particular, they wanted to use this tool when presenting AI-generated outputs to their students. As one participant suggested, ``\emph{there needs to be a very blatant warning signal. I'm thinking about our kids like something that is very bright that grabs attention or like ways that they could report a bias on whatever they're reading.}'' (W1). Participants also described using flags to prompt students to identify problematic outputs themselves. For example, a participant in Workshop 1 requested, ``\emph{a way that [they] could be `like, I think this might be a bias,' and then it could have a category of types of bias, and then the kid could identify theirs.}'' Participants wanted students to actively engage with AI outputs, honing their ability to discern whether outputs are problematic and develop their own point of view rather than passively absorbing information. These findings illustrate how educators view auditing tools as dual-purpose: not only for uncovering problematic outputs but also as a pedagogical aid to build students' critical thinking skills.

Flagging problematic outputs is more technically straightforward than the other features but poses sociotechnical challenges for implementation. Unlike more straightforward hallucination detection in generative AI tools, cultural misrepresentation requires reconciling diverse, subjective judgments. In Hawai`i, many teachers have recently relocated from the mainland, bringing varying levels of cultural knowledge~\cite{civilbeat2019hiring,HIDOE_EmploymentReport_2023-24}. For example, in Workshop 2, when we showed generated images that misrepresented an ``ohe kāpala'', a traditional Hawaiian stamp made from bamboo used to decorate clothing, our participants told us that they had not heard of the artifact before (see Fig.~\ref{fig:probes}). In contrast, when reacting to the same design probes, a participant in Workshop 4 who identifies as Kanaka Maoli (Native Hawaiian) stated: 
\begin{quote}
    \emph{``No, we’re not doing that; we’re culturally grounded. But I can definitely see it how a lot of new teachers in Hawai`i, like, coming from the US would just use [the output], 100\% would just use it without thinking twice.''}
\end{quote}
When educators’ backgrounds do not align with students’ lived experiences, this can result in cultural disconnects or the inadvertent use of educational material that reinforces stereotypes or omits key cultural contexts. These challenges are compounded by uneven levels of cultural and AI literacy among educators, which may hinder their ability to audit AI outputs critically.

\subsection{Design Vignette}
\label{subsec:design_vignette}
\begin{figure*}[t!]
  \centering
\includegraphics[width=\textwidth]{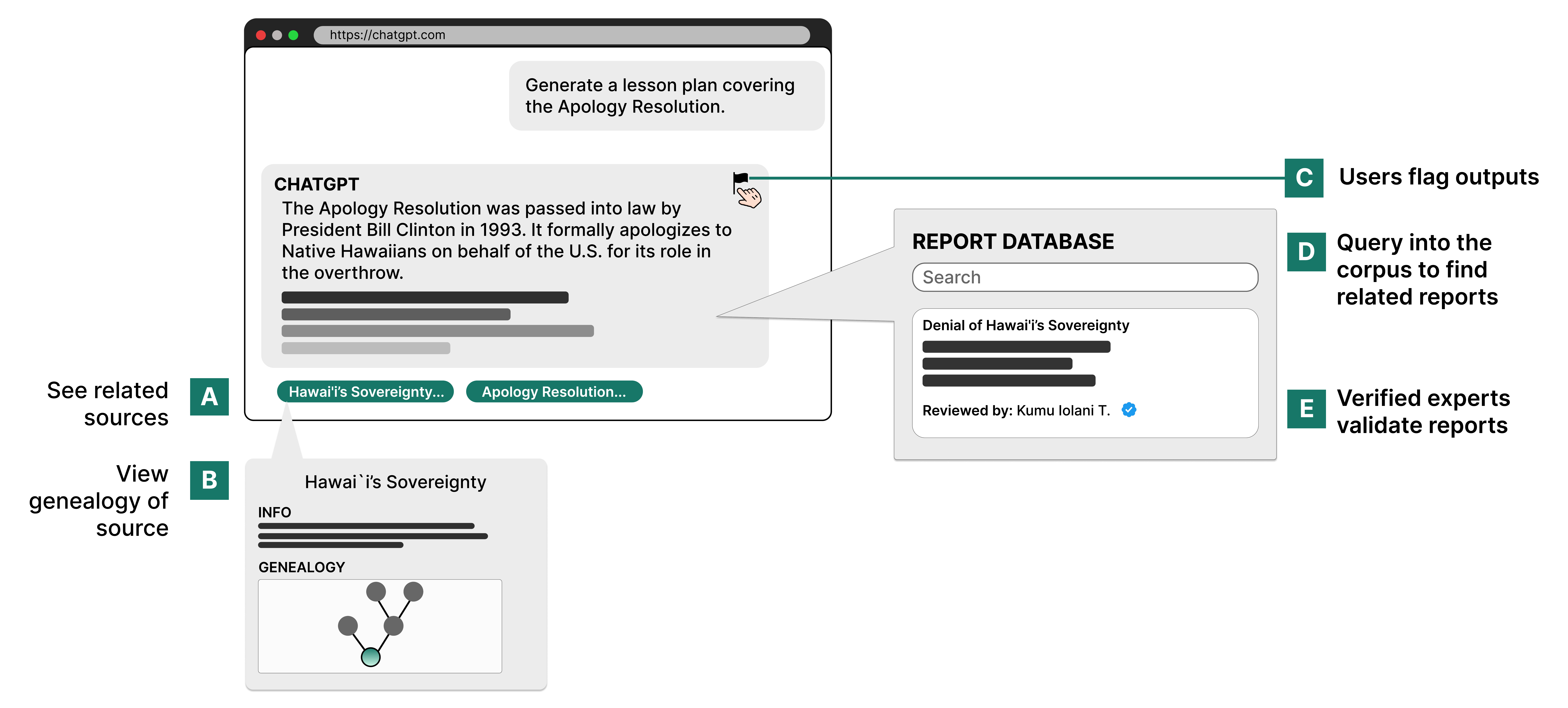}
  \caption{We present an exemplar auditing tool based on the authors distilling the prototyped features, discussions, and context into a cohesive system. The tool is proposed to be built on top of existing AI platforms (e.g., ChatGPT, Gemini). Users can view if there are sources related to the generated output (A) and view more information about who authored the source as well as the genealogy of knowledge (B). If a user is concerned about the output, they can flag it to enter into the report database (C). Newcomers or less experienced auditors can also participate by comparing outputs to preexisting reports in the database (D) and flagging potentially harmful results that are reviewed by more experienced community members (E).}
  \label{fig:design_vignette}
  \Description[Exemplar design of an auditing tool proposed for use alongside generative AI platforms.]{
  Exemplar design of an auditing tool proposed for use alongside generative AI platforms. The interface shows a generated lesson plan about the 1993 U.S. Apology Resolution to Native Hawaiians. Users can interact with multiple auditing features: (A) view related sources connected to the generated text; (B) inspect the genealogy of a given source, including its author and knowledge lineage; (C) flag problematic outputs, which are stored in a shared report database; (D) query the database to find related reports and compare across cases; and (E) see reports validated by verified cultural experts.
  }
\end{figure*}
We conclude by presenting an exemplar system distilled from our workshop findings (Fig.~\ref{fig:design_vignette}). Our design focuses on two core themes that emerged around the importance of knowledge genealogy within Hawaiian epistemology and the diversity of cultural expertise educators have.

\subsubsection{Exemplar System Design} We propose a design that allows educators to trace and contest generative AI outputs in a collective fashion. When educators come across an output, they can inspect relevant sources, which contain attributes about the source's author and their background, allowing users to visualize the genealogy of knowledge. If the user views the output as problematic, they can provide their interpretation on the nature of the harm. These entries are compiled into a shared report database that grows over time. To facilitate knowledge provenance within this report database as well, submissions are linked to prior contributions by the same auditor along with information about that individual and additional trust signals (e.g., response verified by a cultural expert). 

To account for the diversity of expertise, the system scaffolds the development of critical auditing skills. Users who feel uncertain about their interpretation can query into the report database to surface related concerns that others have raised, facilitating learning through doing. In addition, they can flag potentially concerning outputs, which are then routed to a more experienced community member that offers their expertise or validates any interpretation that the user provided. Reflecting participants' desire to use auditing as a educational exercise, students can also engage in flagging outputs or viewing the report database. Importantly, the collected database of flagged outputs and any reference sources should remain locally-hosted and shared only with a trusted community. For one, distrust in large tech corporations or fear of misappropriation may otherwise deter users from contributing. Moreover, as discussed in Sec.~\ref{subsec:sources}, digitizing Indigenous data can become a new form of colonialism if non-community members appropriate cultural resources to train AI systems without any benefit being returned to Indigenous communities \cite{ledward2008hcie, young2019new, Spano_Zhang_2025, young2019new}. 

In the current design, we present the tool as an independent browser extension that is interoperable across multiple generative AI platforms. We note that partnering with existing model or educational technology providers is an alternative choice that could ensure easier maintenance. However, based on survey results from Sec.~\ref{subsec:usage}, teachers' generative AI workflows are distributed across many platforms, requiring auditing tools that provide more flexibility. Again, given the emphasis on ensuring data sovereignty in our workshops, building tools that can be hosted and controlled locally are a top priority.

\subsubsection{Implementation Challenges and Considerations} We conclude by surveying which aspects of our design vignette are feasible in the short term and identify areas where future work can contribute. As discussed in Sec.~\ref{subsec:sources}, source attribution in LLMs remains a significant technical challenge and represents a growing area of research~\cite{khoo2022deepfake, wang2024sourceattributionlargelanguage, tilwani2024neurosymbolicaiapproachattribution}. While reliable attribution may remain difficult, retrieving related sources may be possible using existing NLP and information retrieval techniques~\cite{wang2024sourceattributionlargelanguage}. Beyond technical barriers, however, there are substantial sociotechnical challenges in implementing such a system. A central, and contested, question is who should govern or maintain it. While our findings suggest that control should rest with the community, there was no clear consensus on what ``community governance'' should look like. Feedback from our workshops reflected these tensions: participants in Workshop 4 favored oversight by state education officials (e.g., the Office of Hawaiian Education); Workshop 3 participants preferred avoiding any government or political involvement; and Workshop 1 participants expressed interest in a more decentralized, school-by-school approach.

\section{Discussion}
Our findings show that educators in Hawai`i face competing dynamics around the use of generative AI. On one hand, participants described top-down encouragement from the state to integrate these tools, reflecting an institutional push toward adoption. At the individual level, many workshop participants expressed interest in using generative AI for many reasons, including saving time, creating more localized education content, and offering more interactive classroom activities. At the same time, however, participants voiced a wide range of concerns about the potential harms these technologies may introduce, particularly when applied to tasks related to cultural knowledge. Given that Hawaiian culture is a mandated part of the curriculum, these shortcomings of generative AI are not a hypothetical risk but rather an inevitable problem with which participants need to reckon. These tensions highlight the gap between the push to adopt AI and the lack of safeguards ensuring its appropriateness. While end-user auditing provides a general approach to bridge this gap, our workshops underscore that participants require tools designed with Hawaiian values at the forefront, rather than relying on general-purpose solutions. 

While our study is situated within the cultural and educational context of Hawai`i, we speak to how our findings offer insights that might inform end-user auditing practices for other marginalized communities. Although the specific harms and designs that participants reported are not generalizable, our takeaway that we must consider \emph{community-specific} designs for end-user auditing has broader implications. To date, most work on auditing tools has focused on creating more general-purpose tools that are intended to work across many settings and for many groups of people~\cite{ojewale2025towards,deng2023understanding}. However, as \citet{bird2024must} argued about NLP methods more broadly, this ``one-size-fits-all'' approach problematically treats language as a decontextualized technical issue to solve, stripping away how language is actually used by people within specific communities. This operationalization may not account for the needs, values, or ways of knowing within a specific community, especially those that have been historically marginalized. 

Within HCI, decoloniality has provided a useful framework for understanding how we can center a plurality of perspectives, rather than assuming a universalist default~\cite{wong2020reflections,shahid2023decolonizing,wong2020decolonizing}. In particular, \citet{alvarado2021decolonial} provide an agenda for decolonizing HCI research, listing five pathways (understanding the why, reconsidering the how, changing the for whom, expanding the what, and reflecting on the what for) that can be applied to reorient how we design sociotechnical systems. Our following discussion will touch on three of the pathways in the context of end-user auditing, discussing how to (1) accommodate diverse perspectives when auditing; (2) develop infrastructures that map to community values; and (3) ensure audit outputs benefit the community.

\subsection{Determining who conducts community-centered audits} 
Our findings illustrate that auditing for harms in AI systems requires situated knowledge that individuals external to that community may lack~\cite{qadri2025casethickevaluationscultural,haraway2013situated}. For example, as a participant in Workshop 1 noted when observing a bias probe: ``\emph{we looked at that [generated] picture and we're like that’s definitely not a picture of a Hawaiian classroom, but somebody in Kansas, they might think and say yeah, that's what a classroom looks like because they've never come here.}'' This point echoes calls from recent HCI work advocating for adopting more participatory methods in AI auditing by involving end users, such as the educators from our workshops, as auditors~\cite{deng2025weauditscaffoldinguserauditors,ojewale2025towards}. Moreover, our workshops revealed that within communities, we should not expect participation to be uniform: differences in cultural expertise --- such as between Kānaka or Kama`āina educators and those recently arrived from the mainland --- shaped who could recognize harms. For example, while most participants in our workshops found the design probe related to ``pineapple picking'' to be offensive, some participants in Workshop 2 stated that they ``\textit{don't see it being biased.}'' 

This raises an important question of how to design for the diversity of community members' perspectives and experiences in auditing systems.  As \citet{alvarado2021decolonial} argue, the decolonial path of ``expanding the what'' requires embracing multiple frames of reference and supporting plural knowledge systems. While leveraging users' diverse range of lived expertise is a key benefit for end-user auditing, if applied uncritically, it also creates a veneer of equality as not all community members bring the same forms of knowledge. Rather than treating auditing as a generic task that can be performed by any user, these asymmetries ought to shape whose judgments carry weight when conducting audits. At the same time, it is essential to ensure that the responsibility of auditing does not disproportionately fall on marginalized community members~\cite{smith2011challenging,williams2019uncompensated}. Thus, auditing systems must center the epistemic authority of those most directly connected to the cultural contexts at stake, while still providing pathways for allies to participate.

\paragraph{Design Recommendations} We provide two design recommendations for how community-centered audits can explicitly account for these differences in their design. First, auditing tasks should be designed to accommodate different tiers of involvement or ``trust'' rather than assuming auditors will come in with equal experiences. Following practices from situated learning~\cite{lave1991situated}, those with less experience can start with less involved tasks. After observing experts and demonstrating that they can be trusted with more involved tasks, they can take on more key roles. How these levels of trust are operationalized will depend on the community. Examples could include delineating based on task type (e.g., more interpretive work versus fact-based validation) or issue severity. Second, auditing systems must include mechanisms to handle disagreements and provide pathways for discussion when auditors' judgments conflict~\cite{deng2025weauditscaffoldinguserauditors}. Especially when dealing with AI outputs that do not have a definitive right or wrong answer, such as is in our case when dealing with cultural artifacts, disagreements are both inevitable and informative. Rather than treating these divergences as errors, systems should scaffold structured dialogue~\cite{shaw2025agonistic,deng2025weauditscaffoldinguserauditors}, such as by allowing auditors to document their reasoning, surface alternative interpretations, and view how others within the community have assessed similar cases. This not only supports transparency but also fosters collective sense-making, enabling auditing practices to better reflect the diversity of knowledge present even within communities.

\subsection{Developing infrastructures to support community audits} While efforts to make AI auditing more participatory have largely focused on including end users in evaluating model outputs, we argue for extending this lens to the design of audit tools themselves. The technologies that these audits rely on also imbue their own standards and politics in the same way that individuals' positionality inevitably shapes auditing outputs~\cite{winner1980artifacts}. Relying on one-size-fits-all infrastructures risks imposing universalist notions of what constitutes a harmful or undesired output that may not align with the communities being served. As illustrated in our workshops, points of friction arise when general-purpose tooling fails to accommodate what our participants need to conduct audits. For example, many participants expressed interest in source auditing. Ostensibly, knowledge attribution or retrieval-based systems could provide this functionality; however, conversations with participants revealed that what they needed was not only knowing what the source says but also its genealogy. Since cultural content, such as mo`olelo and proverbs, are open to multiple interpretations, participants emphasized that assessing trustworthiness depends on knowing \emph{whose} interpretation is being surfaced. In contrast, adopting a decolonial approach, instead, asks how we can move from this mindset of universality towards one of pluriversality, ensuring that auditing tools embed the local values of the communities they are intended to serve. This ties to \citet{alvarado2021decolonial}'s pathway of ``reconsidering the how'' which invites us to think about how we can orient our methodologies to match what the communities we are engaging with want or value. In addition, this approach connects to \citet{bird2024must}'s framework for non-extractive NLP, which argues for building language technologies that grant agency to the communities from which they come, by leveraging cultural resources such as elders and engaging the community where they are at.

\paragraph{Design Recommendations} As a concrete recommendation, we propose expanding the frame of participatory auditing: from inviting communities to perform audits toward enabling them to shape the infrastructures that make auditing possible. How do we enact this in practice? In this work, we adopt practices from participatory design (PD) and conduct co-design sessions. However, this method poses its own set of challenges. Prior work in HCI has documented the limitations and even harms of engaging in PD, especially when working with historically marginalized communities~\cite{harrington2019deconstructing}. There are also concerns about how to maintain any tools built going forward, especially if they are created by researchers who may not be incentivized to invest in long-term maintenance~\cite{kotturi2024sustaining}. Expanding on the exercises from our design workshops, we advocate for future work to draw on speculative design approaches, which has proven fruitful for imagining alternative futures particularly for historically marginalized groups~\cite{wong2018speculative,harrington2021eliciting}. In conjunction to extending PD methods to creating auditing tools, we advocate for further explorations into equipping community members with the means to tool build themselves, allowing for community self-determination, rather than relying on researchers or model practitioners as intermediaries. We recognize that building tools requires more technique overhead; however, we see opportunities in leveraging advances in generative AI technologies to lower these barriers. For example, prompting and other lightweight interaction modalities, such as audio-based interfaces, could make it more feasible for people to create or customize auditing tools without requiring deep technical expertise. Already, there are initial explorations of how to support educators in developing their own AI tooling, which can be extended into the auditing context~\cite{harvey2025don}.

\subsection{Reconsidering valued outcomes of AI audits}
Finally, we question \emph{who} benefits from AI audits, mapping to the pathway about changing for whom HCI research is being done~\cite{alvarado2021decolonial}. At the moment, most AI auditing endeavors focuses on offering technical remediation and improving the model output~\cite{saleiro2019aequitasbiasfairnessaudit,raji2020closing, mokander2023auditing}. However, this design centralizes power in the hands of the platform or model providers who can decide whether to address these harms. It also leaves community members uncertain of whether they will benefit from their labor. Furthermore, centering technical solutions presupposes that AI systems ought to be used and simply need remediation, when it is possible that refusal, or choosing not to not use AI altogether, may be the best course of action in the given situation. These considerations leads us to the question ``\emph{who are these audits intended to serve?} \emph{Is the goal of the audit to improve the model's performance, or is it to help the community}? '' In fact, participants in our workshops rarely expressed interest in influencing platforms holistically. Instead, they imagined auditing features that supported locally consequential decisions, such as whether a lesson plan was appropriate for a classroom or whether to use a generative AI system at all.

\paragraph{Design Recommendations} Drawing on \citet{shahid2023decolonizing}’s work on decolonial approaches to content moderation, we adopt their framing that the goal of such practices should be to ``repair, educate, and sustain communities.'' Applying this perspective to AI auditing, we provide design recommendations about how to create tools that can benefit communities directly. An important aspect of this is designing auditing tools that incorporate educational opportunities for community members. Already in our workshops, we saw how participants viewed auditing as a useful educational exercise for their students, encouraging them to engage more deeply with generated outputs rather than taking the content at face value. Lighter-weight interventions may involve reflection prompts that periodically appear as a user is interacting with a generative AI system that asks them to think about what assumptions the model may be making when generating a response. A more heavy-weight approach could be building multi-tiered auditing tools, similar to our exemplar in Sec.~\ref{subsec:design_vignette}, that can support community members in identifying problematic outputs while also scaffolding learning for students or newcomers. Prior work~\cite{solyst2025investigating} has also suggested integrating best practices from learning sciences in the design of auditing tools to help facilitate the process of learning by doing. While there are many design opportunities in this direction, the underlying principle is to build auditing tools that, first and foremost, directly benefit the communities using them.

\section{Limitations and Broader Ethical Reflections}
In this study we sought to make observations and recommendations that could apply generally to practitioners seeking to build AI auditing tools for educators in Hawai`i. To support this, we conducted workshops with teachers located in Hawai`i. Due to the focused nature of our study, the scope of the participant pool is a limitation to our findings. Our workshops were all conducted with elementary-level public school teachers located in O`ahu. Teachers on other islands, private or charter school teachers, and teachers at different grade levels would have different concerns about using AI in educational contexts. In addition, although our sample size is in line with similar works at CHI, the limited participant size (N=22) can hinder generalizability. Furthermore our participants optionally chose to participate in our study after school hours, so the population of participants at each school was self-selecting. This may have led to a participant population that was more familiar or willing to engage with AI systems, whereas teachers who are less familiar or even opposed to AI may have opted not to participate. Finally, we center educators' perspectives in this study; however, there are other stakeholders --- students, administrators, cultural practitioners --- who are not represented in our workshops. Thus, it remains future work to confirm how these observations align with shared values across a broader spectrum of educators in Hawai`i. 

Another limitation of this work is our focus on text-based generative AI systems. The proposed designs in Sec.~\ref{subsec:wireframes} and design vignette are primarily intended for natural language input and output. This decision reflects how participants are primarily using generative AI systems. We expect that the auditing needs, harm types, and required technical solutions will vary across modalities (e.g., for visual or for audio-based systems), necessitating different mechanisms for surfacing harm which are not explored in depth within this work. 

Finally, we want to provide a reflection on the broader ethical considerations of this work. Initially, the missive of this project was to build a general-purpose auditing tool that could be used by educators in Hawai`i. However, after engaging in workshops, it became clear that a generalizable approach to cultural auditing would be detrimental to the community's needs. As described by participants, cultural knowledge is embedded in localized, generational practices, and access to this knowledge requires building deep relationships and trust with community members. Hence, attempts to abstract these culturally significant practices into a generalized tool deployed across communities risks misrepresenting or flattening the epistemologies they aim to preserve. These considerations shifted us away from designing a generalizable tool, and instead ideate on frameworks that support community-centered auditing infrastructures. Ultimately, we argue the decision to build, or not to build, should be made not only considering technical feasibility, but through careful reflection about whether the system can be sustained over time and most importantly whether it provides long-lasting benefits to the community beyond the scope of an academic paper.
\section{Conclusion}
Overall, this work sheds light on educator concerns surrounding AI in the classroom and introduces a framework for co-developing community-centered auditing tools for Indigenous and low-resource contexts. Through four co-design workshops with 22 public school educators in O`ahu, Hawai`i, we identified key worries around AI use in education --- particularly regarding cultural representation --- and surfaced auditing practices educators envision to address these issues. Drawing on these insights, we argue that AI auditing should be framed as a \emph{community practice}: one actively shaped by the people and communities most affected by these technologies. This reframing invites us to reconsider who serves as an auditor within this process, how auditing infrastructures ought to be designed, and what outcomes should be valued in the auditing process.

\begin{acks}
We thank our participants for their contributions, and our elementary school partners for helping facilitate the design workshops. In addition, we also are grateful to the Ulu Lāhui Foundation, Purple Mai`a Foundation, Hawaii Department of Education, the Office of Hawaiian Education, Alika Spahn Naihe, and Amanda Nelson for taking the time to meet with members of our research team. Finally, we are grateful for members of the SALT Lab for their helpful feedback on this work.  

This research was partially supported by the National Science Foundation under award numbers IIS-2247357, CNS-2137784, and CNS-2145584. We would also like to acknowledge support by the Alfred P. Sloan Foundation,  DSO National Laboratories (DSO), VMWare, Google, and Catherine M. and James E. Allchin. Dora Zhao is supported in part by the Paul and Daisy Soros Fellowship for New Americans. Any opinions, findings, conclusions, or recommendations expressed in this material are those of the authors and do not necessarily reflect the views of the National Science Foundation or other supporters. 
\end{acks}

\bibliographystyle{ACM-Reference-Format}
\bibliography{bibliography}

\appendix
\section{Appendix}
\subsection{Study Materials}
We provide the materials used during our co-design workshops including the design exercise prompts and survey questions. 

\subsubsection{Design Exercises} 
\begin{figure*}
    \centering
    \includegraphics[width=\linewidth]{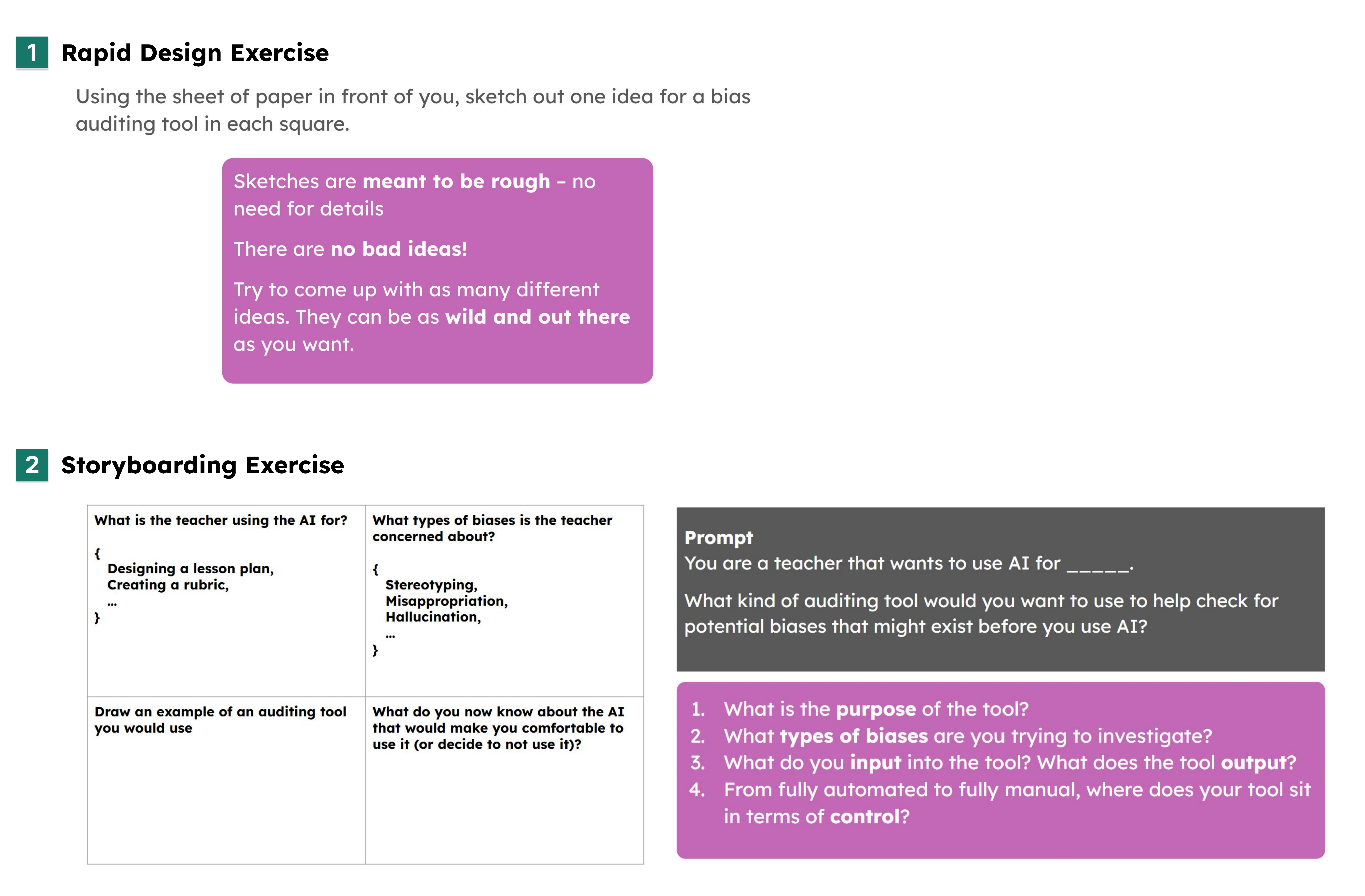}
    \caption{Workshop participants completed one of two design exercises. For participants in Workshops 1 and 2, they took part in a rapid design exercise (top) focused on ideating many divergent designs for auditing tools. For participants in Workshops 3 and 4, in groups of 2-3, they completed a storyboarding exercise for a single design (bottom).}
    \label{fig:design_workshop}
    \Description[Panel A depicts the rapid design exercise, and Panel B shows the storyboarding exercise completed by other participants.]{
    Panel A depicts the rapid design exercise, in which participants sketched multiple rough ideas for AI bias-auditing tools on worksheets. The prompt emphasized exploratory thinking, highlighting that sketches should be rough, that there were no bad ideas, and that participants were encouraged to generate as many creative and unconventional concepts as possible. Panel B shows the storyboarding exercise completed by other participants, who filled out a four-quadrant storyboard describing a single auditing-tool concept.
    }
\end{figure*}

In Fig.~\ref{fig:design_workshop}, we provide the prompts for our two design exercises: rapid prototyping and storyboarding. After the design exercise, participants were invited to share out with the group about designs they created. We asked the following questions to scaffold the sharing process:
\begin{enumerate}
    \item What is the purpose of the tool? What types of biases does it aim to audit? 
    \item How would you use the tool? 
    \item What did you prioritize when designing this tool?
\end{enumerate}

\subsubsection{Survey Questions}
\begin{figure*}
    \centering
    \includegraphics[width=\linewidth]{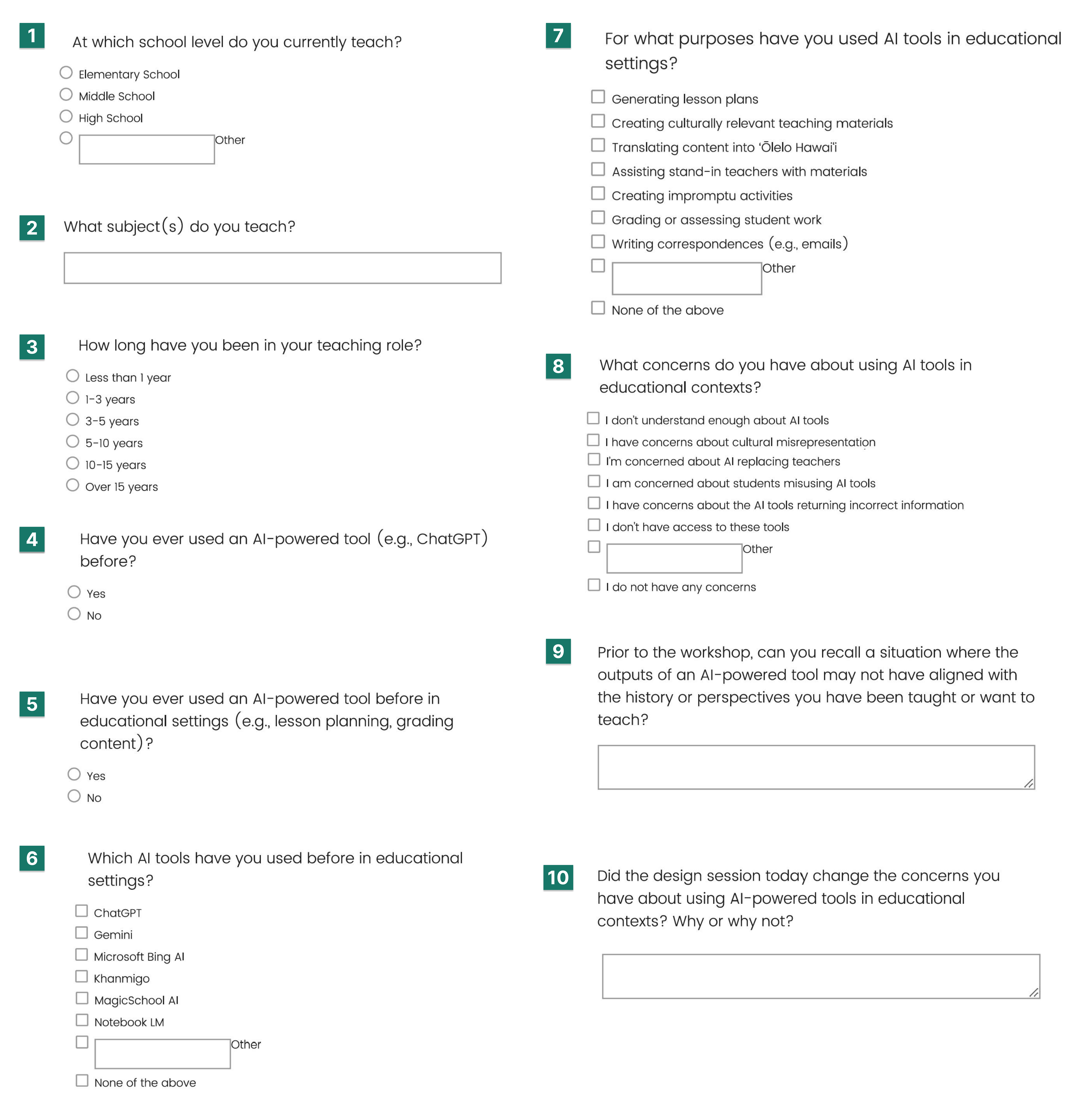}
    \caption{Online Qualtrics survey that participants took after the co-design workshop (Part 1).}
    \label{fig:qualtrics_pt1}
    \Description[Survey questions presented to participants after the workshop.]{Survey questions presented to participants after the workshop, related to their background experience on AI, concerns they had about AI usage in classrooms, and reflections on the workshop.}
\end{figure*}

\begin{figure*}
    \centering
    \includegraphics[width=\linewidth]{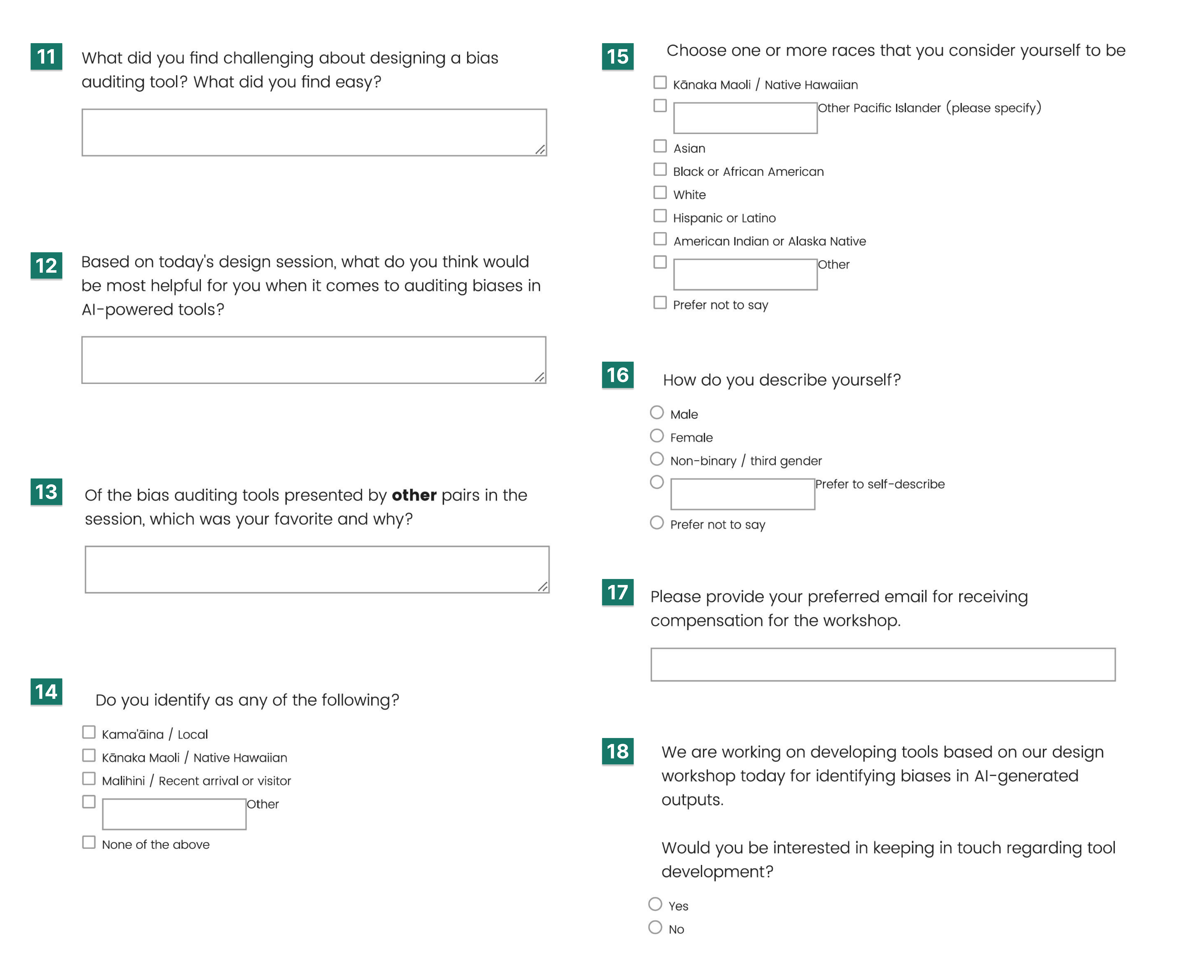}
    \caption{Online Qualtrics survey that participants took after the co-design workshop (Part 2).}
    \label{fig:qualtrics_pt2}
    \Description[Survey questions presented to participants after the workshop.]{Survey questions presented to participants after the workshop, related to their background experience on AI, concerns they had about AI usage in classrooms, and reflections on the workshop.}
\end{figure*}

At the conclusion of the workshop, participants were invited to complete the following survey. Questions are shown in Fig.~\ref{fig:qualtrics_pt1} and~\ref{fig:qualtrics_pt2}.

\subsection{Codebook}
\label{sec:app_codebook}
We provide the codebook produced through our analysis process in Table~\ref{tab:codebook}.
\begin{table*}[h!]
    \centering
    \add{\begin{tabular}{lc}
    \toprule 
    \textbf{Code} & \textbf{Count}\\
    \midrule
    \emph{Concerns about Generative AI} \\
    \midrule 
Coherence (e.g., alignment to Hawaiian culture)&  6\\
Connotations (e.g., Hawaiian culture being treated as part of history) & 6\\
Incorrectness and Hallucinations & 19\\
Irrelevance of Outputs to Hawaiian Culture& 6\\
\quad\quad Only surface level representation of Hawaiian culture & 13\\
Erasure of Cultural Elements & 11\\
Skewedness of Cultural Depiction& 5\\
\quad\quad Not representing the diversity of Hawai`i & 6\\
Specificity of Cultural Elements& 7\\
\midrule 
\emph{Challenges with identifying harms} \\
\midrule 
Disagreements about what is biased & 11\\
Lack of data and sources about Hawaiian culture & 9\\
Need community knowledge to do auditing / identify harms & 20\\
No ground-truth to compare against & 5\\
Lack of self-confidence when identifying bias & 7\\
\midrule 
\emph{Factors influencing design} \\
\midrule 
Bigger picture issues beyond representational bias & 5\\
Grade-level and subject matter influence needs & 4\\
Importance of social support for educators & 6\\
Accounting for the positionality of auditors & 5\\
Being able to trust sources and vet reliability of information & 37\\
Scarcity of data & 4\\
Integrating Hawaiian Values & 4\\
\quad\quad Kilo (learning via observation) & 2\\
\quad\quad Knowledge Provenance & 8\\
\midrule 
\emph{Tensions in design} \\
\midrule 
Automation vs. Reflection & 9\\
Cultural specificity vs. Unawareness & 9\\
Data Sharing vs. Co-option & 3\\
Efficiency vs Quality & 3\\
Stereotyping vs Personalization & 7\\
\midrule 
\emph{Auditing Tool Ideas} \\
\midrule 
Perspective auditing & 14\\
Auditing the prompt or person, not the output & 2\\
Comparing across models and outputs & 9\\
Source auditing & 22\\
Flagging bias & 3\\
\bottomrule
    \end{tabular}}
    \caption{Codebook with themes and sub-codes surfaced through reflexive thematic analysis described in Sec.~\ref{subsec:analysis}}.
    \label{tab:codebook}
    \Description[List of codes used during the analysis process.]{
    List of codes used during the analysis process. The five parent codes are as follows: concerns about generative AI, challenges with identifying harms, factors influencing auditing tool design, tensions in auditing tool design, and auditing tool ideas.
    }
\end{table*}

\end{document}